\documentclass[aps,10pt
]{revtex4}
\def\theequation{\arabic{section}.\arabic{equation}}
\usepackage{psfrag}
\usepackage{color}
\usepackage{mathrsfs}
\usepackage{graphicx}
\usepackage[colorlinks=true,linkcolor=blue,citecolor=magenta,urlcolor=blue]{hyperref}
\usepackage{amssymb, bm}
\usepackage{amsmath, amsthm}
\usepackage{epstopdf}
\usepackage{hyperref}
\usepackage{enumerate}
\usepackage{longtable}
\usepackage{float}
\usepackage{array, multirow}
\usepackage{fancyhdr}
\usepackage{amsfonts} 
\usepackage{array}

\newtheorem{proposition}{Proposition}
\theoremstyle{remark}
\newtheorem{remark}{Remark}

\newcolumntype{C}[1]{>{\centering\arraybackslash}p{#1}} 

\newcommand{\dd}{\mathrm{d}}

\renewcommand\thesection{\arabic{section}}

\begin{document}	
	\title{Odd-even dichotomy in the waveforms of modified Emden oscillators of arbitrary integer power} 
	
	\author{J. de la Cruz} 
	\email{josue.delacruz@ipicyt.edu.mx; ORCID: 0000-0001-5943-5752}
     \author{H.C. Rosu}
    \email{hcr@ipicyt.edu.mx; ORCID: 0000-0001-5909-1945}
	\affiliation{Instituto Potosino de Investigaci\'on Cient\'{\i}fica y Tecnol\'ogica,\\ 
		Camino a la Presa San Jos\'e 2055, Colonia Lomas 4a Secci\'on, 
		78216 San Luis Potos\'{\i}, S.L.P., M\'exico}
	
	\bigskip
	\bigskip
  \begin{abstract}
Closed-form waveforms of the modified Emden oscillators $\ddot x+(q+2)kx^q\dot x+k^2x^{2q+1}+\omega^2x=0$ of arbitrary natural power $q$, and of their damped $c$-deformed versions, are obtained through a generalized commutative factorization that reduces them to Bernoulli equations. We show that commutative factorizations of Li\'enard equations are gauge-conjugated forms of the nonlocal linearizing transformation of Chandrasekar, Senthilvelan, Kundu and Lakshmanan, the factorization constant being half the difference of the roots of the associated linear oscillator. These oscillators display an odd--even dichotomy: for odd $q$ they are isochronous, with period $2\pi/\omega$ independent of the amplitude, within an explicitly bounded range of the integration constant, while for even $q$ the oscillations decay algebraically, as $t^{-1/q}$. The dichotomy is confirmed by an elliptic polar representation, in which the phase rotates uniformly for every $q$, and, for even $q$, by the Bendixson--Dulac criterion applied to the divergence of the vector field. We present a dissipative Lagrangian of Lurie--Rayleigh type and, for $q=1$, an autonomous Sundman linearization that yields a rational first integral, a Jacobi last multiplier and a non-dissipative Lagrangian; for every $q$, $\omega^2x^2+(\dot x+kx^{q+1})^2$ is a Darboux polynomial providing a global Jacobi last multiplier. Explicit waveforms and phase portraits are given for $q=1,\dots,4$, with and without damping, and a recent application to magnon dynamics in the magnetic semiconductor CrSBr is discussed.

		\medskip
		
		
	\end{abstract}

	\maketitle
\numberwithin{equation}{section}
\renewcommand\thesection{\arabic{section}}
	\section{Introduction}\label{sec1}
The modified Emden class of nonlinear oscillators of arbitrary positive integer power $q$ is defined by the following differential equation
	\begin{equation}\label{eq0}
		\ddot{x} +(q+2)kx^q\dot{x}+k^2x^{2q+1}+\omega^2 x=0~,
	\end{equation}
	which is a Li\'enard nonlinear differential equation 
	\begin{equation}\label{eq1}
		\ddot{x} +f(x)\dot{x} + g(x) = 0~,
	\end{equation}
with $f(x)=(q+2)kx^q$ and $g(x)=k^2x^{2q+1}+\omega^2 x$. We consider $k$ and $\omega^2$ as arbitrary real constants. For $k=0$, this equation reduces to the pure harmonic oscillator.
	
The case $q=1$ of \eqref{eq0} was first considered by Chandrasekar {\em et al.} \cite{CSL2005} in the form 
	\begin{equation}\label{eq01}
\ddot{x} +\kappa x\dot{x}+\frac{\kappa^2}{9}x^3+\lambda x=0~, \quad \kappa=3k~,
	\end{equation}
which we call the $\lambda$-modified Emden equation; it allows one to include the hyperbolic case, $\lambda<0$, and the pure nonlinear case, $\lambda=0$, as well. The general form of the latter case, $\ddot x+\alpha x\dot x+\beta x^3=0$ was shown by the same group \cite{CSL2007} to be integrable, explicitly or by quadrature, for all $\alpha$ and $\beta$, with time-independent Hamiltonians over the full parameter range.
For the oscillator case, $\lambda\equiv \omega^2>0$, they noticed the intriguing properties that the frequency is independent of amplitude, just as in the case of linear harmonic oscillators, 
and that it has conservative Hamiltonian behavior despite the presence of the nonlinear, position-dependent damping term $\kappa x\dot x$, whose sign changes with $x$.
In the seminal paper \cite{CSKL2006} (also abbreviated as CSKL in the following), they developed a systematic method based on a nonlocal transformation connecting certain families of nonlinear oscillators to linear ones, including \eqref{eq0} and its $c$-deformed version
	\begin{equation}\label{eq:underdamped_lienard}
		\ddot{x} + \bigl(2c + (q+2)k x^q\bigr)\dot{x} + k^2x^{2q+1} + 2kcx^{q+1} + (\omega^2 + c^2)x = 0~,
	\end{equation}
which for $k=0$ reduces to the damped linear oscillator $\ddot{x} + 2c\dot{x}+ (\omega^2 + c^2)x = 0$. The nonlocal transformation of CSKL \cite{CSKL2006} linearizes \eqref{eq0} and \eqref{eq:underdamped_lienard} for arbitrary $q$ and gives their general solution in integral form.

	Further, Iacono and Russo \cite{IR2011} studied the odd-$q$ part of these oscillators 
by an approach involving an integrability condition due to Sabatini \cite{Sab1999}. 
Factorizations reducing Li\'enard-type equations, including coupled modified Emden equations, to Bernoulli equations were used by Hazra \emph{et al.} \cite{Hazra2012}, and in our previous work \cite{Ediesca2025} we obtained the isochronous odd-$q$ solutions through commutative factorization.
Chandrasekar {\em et al.} \cite{C2012} presented a complete analysis of coupled cubic oscillators ($q=1$) of this kind. 
The same group also studied coupled oscillators of this type based on an approach in which the integrals of motion are in the same form both for the linear and the nonlinear oscillators \cite{RamyaP2022}.

Very recently, the case $q=1$ of (\ref{eq0}) has been derived from an equation of Levinson-Smith type and also its bi-Hamiltonian character has been established in \cite{bagchi2025}.
It is also worth mentioning the work of Mustafa \cite{Mustafa2023}, who derived the mass-dependent modified Emden equation from the Euler-Lagrange equations for a non-conservative quartic anharmonic potential.

	Here, we study these oscillators using the generalized commutative factorization approach and the associated Bernoulli-type equation, following the method we developed in \cite{pla2025,Ediesca2025}, recovering, for arbitrary $q$, the general solutions obtained in \cite{CSKL2006}. Proposition~\ref{prop:commfact} below shows that the factorization method is an equivalent, more algebraic route to the nonlocal transformation of \cite{CSKL2006}.
Furthermore, we discuss the dissipative Lagrangian of Lurie--Rayleigh type of these oscillators, the Bendixson-Dulac criterion, and their polar representation. 
The following results with various degrees of novelty are reported:
\begin{enumerate}
\item an if-and-only-if characterization (Proposition~\ref{prop:commfact}): a Li\'enard equation admits a commutative factorization exactly when it is a gauge-conjugated image of a constant-coefficient linear oscillator under the CSKL map, the factorization constant being half the difference of the linear roots;
\item a unified treatment of odd and even $q$, with explicit harmonic-sum forms of the waveforms, a closed-form bound for the singular range of the integration constant, $|c_1|\le q\frac{(2n)!!}{(2n+1)!!}\frac{k}{\omega}\sim\sqrt{\pi q/2}\,\frac{k}{\omega}$, and the algebraic envelope $t^{-1/q}$ for even $q$;
\item an elliptic polar representation for arbitrary $k$, $\omega$ and $c$, in which the phase rotates uniformly, $\dot\theta_c=-\omega$, also in the damped case;
\item explicit damped waveforms of \eqref{eq:underdamped_lienard} for $q=1,\dots,4$ and a three-regime classification ($c=0$, $c>0$, $c<0$) according to the parity of $q$;
\item a Lurie--Rayleigh dissipative Lagrangian for every $q$, whose sign structure reflects the odd--even dichotomy;
\item for every $q$, the Darboux polynomial $\omega^2x^2+(\dot x+kx^{q+1})^2$ and the corresponding global Jacobi last multiplier, a rational global first integral for odd $q$, and the absence of a single-valued one for even $q$; for $q=1$, an autonomous Sundman linearization yielding a rational first integral;
\item the connection with the experimental modelling of magnon dynamics in CrSBr \cite{Brennan2026}.
\end{enumerate}

The organization of this paper is the following.
In Section \ref{sec2}, we shortly review the commutative factorization method that allows integrability through the connection with Bernoulli equations. In Section \ref{sec3}, the commutative factorization is shown to be equivalent to a gauge-conjugated form of the CSKL nonlocal transformation method. In the next Section \ref{sec4},
we apply the method to the modified Emden oscillators. The dissipative Lagrangian in Lurie's formalism \cite{Lurie} is presented in Section \ref{sec5}, followed by a brief discussion of the Bendixson-Dulac 
criterion and the polar representation in Sections \ref{sec6} and \ref{sec7}, respectively. Section \ref{sec8} is dedicated to first integrals, Jacobi last multipliers and Lagrangians for the autonomous modified Emden equations, where we follow the approach through autonomous Sundman nonlocal transformations of Sinelshchikov \cite{Sin2020} for $q=1$ and use Darboux polynomials for general $q$.
In section \ref{sec9}, we deal with the case corresponding to \eqref{eq:underdamped_lienard} when the factorization functions differ by a complex constant instead of a pure imaginary one, including the regime $c<0$. Possible applications of these equations are commented in Section \ref{sec10} and the conclusions are in Section \ref{sec11}. There is also an Appendix in which the connection with a Gin\'e--Sinelshchikov class of nonlinear oscillators \cite{GS2024} is presented.
\numberwithin{equation}{section}                  %
\renewcommand\thesection{\arabic{section}}        %
\section{From nonlinear factorization to Bernoulli equation}\label{sec2} 
In \cite{pla2025}, we factored three types of Li\'enard equations, including the $\omega^2=0$, $q=1$ case of \eqref{eq0}, into two first-order differential operators 
and reduced them to equivalent Bernoulli differential equations. The factorization
	\begin{equation}\label{eq2}
		[ D-\phi_2(x) ] [ D-\phi_1(x) ]x=0~, \quad D=\frac{d}{dt}~
	\end{equation}
	is achieved if the factorization functions $\phi_i$ satisfy the conditions \cite{rcp1,rcp2}
	\begin{eqnarray}
		&\qquad \quad \phi_2+\frac{d(\phi_1x)}{dx}=-f(x)~, \label{eq3}\\
		&\phi_1 \phi_2 x =g(x)~.\label{eq3bis}
	\end{eqnarray}
Equation (\ref{eq3bis}) shows that these factorizations can be efficient tools in solution search especially when $g(x)/x$ is a polynomial since $\phi_1$ and $\phi_2$ can be chosen from its factoring polynomials. 
	The commutation case occurs when the factorization functions are of the form $\phi_1=\phi+\mathsf c$ and $\phi_2=\phi-\mathsf c$, which allows the reverting of the factorization brackets without changing the equation \cite{Springer2024}.
	In the recent paper \cite{pla2025}, we showed that a slightly generalized form of the commutative factorization described in \cite{Springer2024} provides the general solutions of (\ref{eq0})
	as solutions of an associated Bernoulli equation, and the case $q=1$ has 
	been briefly presented in this approach.
 Using the intermediate function $\Phi(x,t)$ defined by $\Phi(x,t)=[D-\phi_1(x)]x$ turns (\ref{eq2}) into the system  
	\begin{align}
		&  \dot{x}-\phi_1(x)x=\Phi(x,t)~, \label{eq5}\\
		&\dot{\Phi}-\phi_2(x)\Phi =0~, \label{eq4}
	\end{align}
which can be rewritten as the quasi-linear first-order partial differential equation
	\begin{equation}
\frac{\partial\Phi}{\partial x} \left( \Phi + x\phi_1  \right)+ \frac{\partial\Phi}{\partial t} = \phi_2 \Phi, \label{eq6}
	\end{equation}
that can be solved by proposing the {\em ansatz} $\Phi(x,t)=x\zeta(t)$. 
Then, for the function $\zeta$, one obtains the first-order differential equation
	\begin{equation}\label{eq7}
		\frac{d\zeta(t)}{dt} + \zeta^2(t) = (\phi_2-\phi_1) \zeta(t).
	\end{equation}
	If the commutative factorization condition is considered, then the subtraction of factorization functions $\phi_2 -\phi_1 = -2\mathsf c$ holds
	for $\mathsf c\equiv$ const., and \eqref{eq7} reduces to
	\begin{equation} \label{eq8}
		\frac{d\zeta(t)}{dt} + 2\mathsf c \zeta(t)=- \zeta^2(t),
	\end{equation}
	which is a Bernoulli differential equation of nonlinear order two. Taking $\mathsf c =i\tilde{c}$, with $\tilde{c}\in \mathbb{R}$, one obtains the following solution of \eqref{eq8}:
	\begin{eqnarray}\label{eq9c}
		\zeta(t; \delta) &=& -\tilde{c}  \left[i+\tan( \tilde{c} t+\delta)\right] 
	\end{eqnarray}
where $\delta$ is an integration constant.
	This last result allows us to rewrite the first-order ODE \eqref{eq5} in the form
	\begin{equation}\label{eq10}
		\dot{x}-\zeta(t; \delta)x=\phi_1(x)x~,
	\end{equation}
	namely as a Bernoulli differential equation when $\phi_1(x)$ is a monomial function with the nonlinearity of one order higher than the order of the $\phi$'s.
	Its solution provides the general solution of \eqref{eq1} factored in the (commutative) form given in \eqref{eq2}.

    In the following, we shall use the symmetric form of the commutative factorization of \eqref{eq1} which is more convenient for calculations
	\begin{equation}
		[D - \phi(x) +\mathsf c ] [ D- \phi(x) +\mathsf c^*]\, x=0~, \label{eq11}
	\end{equation}
	where $\mathsf c^*$ is the complex conjugate of $\mathsf c$. This symmetric form agrees with $\phi_{1,2}=\phi\pm \mathsf c$ only when $\mathsf c$ is purely imaginary since then $\mathsf c^*=-\mathsf c$. The sans-serif $\mathsf c$ denotes the factorization constant throughout, while $c$ is reserved for the damping constant of \eqref{eq:underdamped_lienard}.
	
	From \eqref{eq11} the following compatible first-order ODE is obtained
	\begin{equation}\label{eq12}
		\dot{x}- (\phi(x) +\mathsf c)\,x=\zeta(t)x~,
	\end{equation}
	and by using \eqref{eq9c}, we obtain for this case the following differential equation
	\begin{equation}\label{eq13} 
		\dot{x}+ \tilde{c} \tan(\tilde{c} t+\delta)\, x=\phi(x) x~,
	\end{equation}    
which is an integrable Bernoulli differential equation if $\phi(x)$ is a monomial or a monomial plus a constant with the same solutions as the nonlinear second-order differential equation. 

\numberwithin{equation}{section}
\renewcommand\thesection{\arabic{section}}
\section{Commutative factorization as a gauge-conjugated CSKL transformation}\label{sec3}  

We write the nonlocal CSKL transformation, equation (12) of CSKL, as $U=x^{\mathfrak n}e^{\Psi}$, where $\Psi(t)=\int^t\big(\beta x^{\mathfrak m}+\gamma\big)\,dt'$ is evaluated along the solution $x(t)$, so that $\dot\Psi=\beta x^\frak{m}+\gamma$; here and below, equation numbers (7), (8), (11), (12), (13) and (55) quoted with the label CSKL refer to \cite{CSKL2006}. 
For any constant $r$,
\begin{equation}\label{a1}
(D-r)\big(e^{\Psi}x^\frak{n}\big)=e^{\Psi}(D+\dot\Psi-r)\,x^\frak{n}~. 
\end{equation}
Applying this twice to $\ddot U+\frak{c}_1\dot U+\frak{c}_2U=(D-m_2)(D-m_1)U=0$, where $m_{1,2}$ are the roots of $\tilde{\lambda}^2+\frak{c}_1\tilde{\lambda}+\frak{c}_2=0$, gives
\begin{equation}\label{a2}
\big[D+\beta x^\frak{m}+\gamma-m_2\big]\big[D+\beta x^\frak{m}+\gamma-m_1\big]\,x^\frak{n}=0  
\end{equation}
after division by $e^\Psi$.
This is the factored form of the operator in  CSKL (55). These factors differ by the constant $m_1-m_2$, so they commute. Every CSKL equation therefore carries a commutative factorization acting on $x^\frak{n}$.

For $\ddot x+f(x)\dot x+g(x)=0$, the factorization $[D-\phi_2][D-\phi_1]x=0$ with $\phi_{1,2}=\phi\pm\mathsf c$ (constant $\mathsf c$) requires
\begin{equation}\label{a3}
f=-(2\phi+x\phi'),\qquad g=(\phi^2-\mathsf c^2)\,x .
\end{equation}
Comparison of \eqref{a2} with \eqref{a3} yields the following conditions on $U$:
\begin{enumerate}
\item $\frak{n}=1$. Otherwise the term $(\frak{n}-1)\dot x^2/x$ of CSKL (13) takes the equation out of Li\'enard form. General $\frak{n}$ reduces to $\frak{n}=1$ under the point map $z=x^\frak{n}$, with exponent $q=\frak{m}/\frak{n}$.
\item $\beta=k$ and $\gamma$ constant, so that $\phi$ depends on $x$ alone. With $\beta(t)$, $\gamma(t)$ the factorization remains commutative, but its factorization functions depend on $t$.
\item $\frak{m}=q$ arbitrary. Then
\begin{equation}
\phi=-kx^q-\gamma-\frac{\frak{c}_1}{2},\qquad \phi_1=-kx^q-\gamma+m_1,\qquad \phi_2=-kx^q-\gamma+m_2,
\end{equation}
and indeed $-(2\phi+x\phi')=(q+2)kx^q+\frak{c}_1+2\gamma$ and $(\phi^2-\mathsf c^2)x$ reproduce CSKL (13) with $\frak{n}=1$.
\item The factorization constant is half the difference of the linear roots,
\begin{equation}
\mathsf c=\frac{m_1-m_2}{2}=\frac12\sqrt{\frak{c}_1^2-4\frak{c}_2},
\end{equation}
independent of $\gamma$. An overdamped linear equation gives real $\phi$'s; an oscillatory one gives imaginary $\mathsf c=i\omega$, the choice $\tilde c=\omega$ of~\cite{Ediesca2025}; critical damping $m_1=m_2$ gives $\phi_1=\phi_2$, a perfect-square factorization, for which \eqref{eq7} 
becomes $\dot\zeta+\zeta^2=0$, $\zeta=1/(t+C)$, and the tangent solutions degenerate.
\item $\gamma$ is pure gauge. The nonlinear equation depends only on $\frak{c}_1+2\gamma$ and $\frak{c}_1^2/4-\frak{c}_2$. Replacing $U$ by $e^{st}U$ shifts $(\gamma,m_1,m_2)\to(\gamma+s,m_1+s,m_2+s)$ and leaves the equation unchanged. The choice $s=-m_1$ puts one root at zero, so the linear equation becomes first order in $\dot U$. 
    For equation \eqref{eq0}, the complex gauge is $-i\omega$ and for equation \eqref{eq:underdamped_lienard} it is $c-i\omega$.
\item Variable-coefficient linear equations do not add anything new. A commutative factorization $(D-r_2(t))(D-r_1(t))$ with $r_1-r_2$ constant gives $\ddot U-2r\dot U+(r^2-\mathsf c^2-\dot r)U=0$, which $U=e^{\int r\,\dd t}\tilde U$ turns into $\ddot{\tilde U}=\mathsf c^2\tilde U$.
\end{enumerate}

\begin{proposition}\label{prop:commfact}
A Li\'enard equation $\ddot x+f(x)\dot x+g(x)=0$ admits a commutative factorization with $\phi_{1,2}=\phi(x)\pm\mathsf c$ if and only if it is the image of a constant-coefficient linear oscillator $\ddot U+\frak{c}_1\dot U+\frak{c}_2U=0$, with $\frak{c}_1^2/4-\frak{c}_2=\mathsf c^2$, under the general CSKL map (7) (with their $f$ renamed $F$), $U=x\,e^{\int F(x)\,\dd t}$ with $F=-\phi-\frak{c}_1/2$, up to a gauge factor $e^{st}$.  
\end{proposition}
\begin{proof}
With $F=-\phi-\frak{c}_1/2$, one has $-(2\phi+x\phi')=2F+xF'+\frak{c}_1$ and $(\phi^2-\mathsf c^2)x=(F^2+\frak{c}_1F+\frak{c}_2)x$, which is CSKL equation~(8). The converse follows from \eqref{a2} with $\mathfrak n=1$ and $\dot\Psi=F(x)$, i.e. $U=x\,e^{\int F\,dt}$.        
\end{proof}

The CSKL monomial map (12) with $\frak{n}=1$ is the subcase $\phi=-kx^q-\gamma-\frak{c}_1/2$. Non-monomial factorizations correspond to other choices of $F$ in CSKL (7); for example, CSKL (11), with $F=k/x^2$, is also commutatively factorizable.

\medskip

Thus, we have the following equivalences between the commutative factorization and the CSKL nonlocal transformation method (taking $\mathfrak n=1$, $\beta=k$ and $\mathfrak m=q$, so that $\dot\Psi=kx^q+\gamma$; $\Phi$ is the intermediate function of \eqref{eq5}):
\begin{center}
\small
\begin{tabular}{@{}ll@{}}
Commutative factorization & \qquad CSKL\\
$\phi_{1,2}$ & \qquad $-kx^q-\gamma+m_{1,2}$\\
$\mathsf c$ ($i\tilde c$) &\qquad  $(m_1-m_2)/2$\\
$\Phi
=[D-\phi_1]x=x\zeta$ & \qquad $e^{-\Psi}(D-m_1)U$\\
$\zeta$ & \qquad $\dot U/U-m_1$\\  
$\dot\zeta+\zeta^2=(\phi_2-\phi_1)\zeta$ & \qquad $\dot w=-(w-m_1)(w-m_2)$, $w=\dot U/U$\\
$\dot x-\zeta x=\phi_1x$ & 
\qquad $\dot x=(\hat a-\gamma)x-kx^{q+1}$, $\hat a=\dot U/U$ for given $U$\\
\end{tabular}
\end{center}
The third row follows from \eqref{a1}: $e^{-\Psi}(D-m_1)\big(e^{\Psi}x\big)=(D+\dot\Psi-m_1)\,x=[D-\phi_1]\,x$, and dividing by $x$ gives the fourth row, $\zeta=\dot U/U-m_1$, since $U=x\,e^{\Psi}$.
For equation \eqref{eq0}: $\frak{c}_1=0$, $\frak{c}_2=\omega^2$, $\gamma=0$, $m_{1,2}=\pm i\omega$, giving the factors $D+kx^q\mp i\omega$. For equation \eqref{eq:underdamped_lienard}: $\frak{c}_1=2c$, $\frak{c}_2=\omega^2+c^2$, $\gamma=0$, $m_{1,2}=-c\pm i\omega$, giving $A,A^\dagger=D+kx^q+c\pm i\omega$. 
In the fifth row, the logarithmic derivative $w$ of the linearizing variable $U$ is $w=\dot x/x+kx^q+\gamma$ for the monomial map, 
and it satisfies the Riccati equation $\dot w=-(w-m_1)(w-m_2)$; we call it the Riccati variable. It is not a first integral because it changes along the trajectories, and conserved quantities arise only after its Riccati equation has been solved. The commutative factorization amounts to the shift $\zeta=w-m_1$, which converts this Riccati equation into the Bernoulli equation \eqref{eq8}, since $\dot\zeta+\zeta^2=(m_2-m_1)\zeta=-2\mathsf c\,\zeta$. For $\gamma=0$, the Iacono--Russo variable used in Section \ref{sec7}  
is $y=\dot x+kx^{q+1}=x\,w$. 

\numberwithin{equation}{section}
\renewcommand\thesection{\arabic{section}}
	\section{Commutative Factorization of Modified Emden Oscillators}\label{sec4}  
The modified Emden oscillator equation has factorization functions of the commutative type $\phi_1=-kx^q+i\omega$, $\phi_2=-kx^q-i\omega$ ($\phi=-kx^q$, $\mathsf c=i\omega$, $\tilde{c}=\omega$)
\begin{equation}
	\left( \frac{d}{dt} + k x^q + i\omega \right)\left( \frac{d}{dt} + k x^q - i\omega \right) x = 0~.
\end{equation}
 
The equivalent Bernoulli equation reads
\begin{equation}\label{qB}
	\dot{x}+\omega\tan(\omega t+\delta)x=-kx^{q+1}~.
\end{equation}
The change of the dependent variable $x=u^{-1/q}$ turns (\ref{qB}) into the linear equation
\begin{equation}
	\dot{u}-q\omega\tan(\omega t+\delta)u=qk~,
\end{equation}
which leads straightforwardly to the $x$ solutions
\begin{equation}\label{general_q}
	x(t)=\frac{\cos(\omega t+\delta)}{\left(c_1+qk\int \cos^{q}(\omega t+\delta)dt\right)^{1/q}}~,
\end{equation}

Taking $s=\omega t+\delta$ and $v = \sin s$, the integral in the denominator can be put in the form
\begin{equation}
	\frac{1}{\omega}\int \cos^{q}(s) ds=\frac{1}{\omega}\int (1 - v^2)^{\frac{q-1}{2}} dv~,
\end{equation} 
which has the form of the binomial integral by identifying ${\rm m}=0,a=1,b=-1,{\rm n}=2,p=\frac{q-1}{2}$:
\begin{equation}
	\int\cos^q(\omega t+\delta)dt=\frac{1}{\omega}\int(1-v^{\rm n})^pdv~.
\end{equation}
It is well known that these integrals can be reduced to simple forms if at least one of the following conditions is satisfied \cite{Nikolsky}:
 $\{p,\frac{{\rm m}+1}{{\rm n}},p+\frac{{\rm m}+1}{{\rm n}}\}\in\mathbb{Z}$.
In our case, the relevant conditions are $p\in\mathbb{Z}$, which holds for odd $q=2n+1$, and $p+\frac{1}{2}\in\mathbb{Z}$, which holds for even $q=2n$,
respectively \cite{Abramowitz}:
	\begin{align}
		&x_o(t)=\frac{\cos(\omega t+\delta)}
{\sqrt[2n+1]{c_1+\frac{2n+1}{2^{2n}}\frac{k}{\omega}\sum_{j=0}^n\binom{2n+1}{j}\frac{\sin((2n+1-2j)(\omega t+\delta))}{2n+1-2j}}}~,
\label{odd}\\
		&x_e(t)=\frac{\cos(\omega t+\delta)}
		{\sqrt[2n]{c_1+\frac{2n}{2^{2n}}\frac{k}{\omega}\left(\binom{2n}{n}\omega t+ 2\sum_{j=0}^{n-1}\binom{2n}{j}\frac{\sin((2n-2j)(\omega t+\delta))}{2n-2j}
\right)}}~.
\label{even}
	\end{align}
	
	In the odd cases (\ref{odd}), only trigonometric functions appear. This leads to periodic solutions and phase portraits with closed orbits. The period for these functions is $T=\frac{2\pi}{\omega}$ for all values of $c_1$ or $k$. This guarantees that the family of modified Emden oscillators is isochronous for odd values of $q$. 
	
	To prove the period value for the odd cases, we apply a translation operator to the function $x_o(t)$: $\hat{T}x_o(t)=x_o(t+T)$. Taking this displacement and using some trigonometric identities, we find that the value of $T$ is the same for all odd $q$ and is independent of the value of $c_1$.
	
	Due to the rational form of \eqref{odd}, a singular region exists for certain combinations of $c_1$ and the equation parameters $(k,\omega)$. This region  occurs when the initial condition takes values in:
{\small	\begin{equation}\label{eq3.44}
		|c_1|\leq\frac{(2n+1)k}{2^{2n}}\max\left[t\sum_{j=0}^n\binom{2n+1}{j}{\rm sinc}((2n-2j+1)\omega t)\right],
	\end{equation}}
	where ${\rm sinc}(x)=\sin (x)/x$. 
	
	A compact formula for the singular region follows directly. With $v=\sin(\omega t+\delta)$, the denominator of \eqref{general_q} is $c_1+\frac{qk}{\omega}\int_0^{v}(1-s^2)^n\,ds$, whose extreme values over a period are $\pm\frac{qk}{\omega}\int_0^1(1-s^2)^n\,ds=\pm\frac{qk}{\omega}\frac{(2n)!!}{(2n+1)!!}$ (Wallis integral). Hence the solution is singular when
		\begin{equation}\label{eq3.44b}
			|c_1|\leq q\,\frac{(2n)!!}{(2n+1)!!}\,\frac{k}{\omega}\equiv 2^n\frac{n!(2n+1)}{(2n+1)!!}\frac{k}{\omega}\;\sim\;\sqrt{\frac{\pi q}{2}}\,\frac{k}{\omega}~,~~q=2n+1~,
		\end{equation} 
	
	For the even cases (\ref{even}), a linear/secular function of $t$ occurs in the denominator. For large values of time, $x(t)$ goes to zero asymptotically,
	which generates stable trajectories in the phase portrait.

	Taking the asymptotic behavior for the even solutions we find that the envelope function $(E_{2n})$ depends on the parameters $k$, $q=2n$ and the initial condition $c_1$ as:
\begin{equation}\label{env}
E_{2n}(t; k,c_1)\approx\pm\left(\frac{2n}{2^{2n}}\binom{2n}{n}kt+c_1\right)^{-1/2n}~,
\end{equation}
where $\binom{2n}{n}$ is the binomial coefficient.

\medskip

	In the remainder of this section, we will illustrate the first two odd and two even cases, $q\in \{1,2,3,4\}$. The arbitrary phase $\delta$ is set to zero to write more compact formulas.

\medskip

{\bf The odd cases $q=1,3$}\\
	
\begin{figure}[H]\centering
        \includegraphics[width=0.85\linewidth]{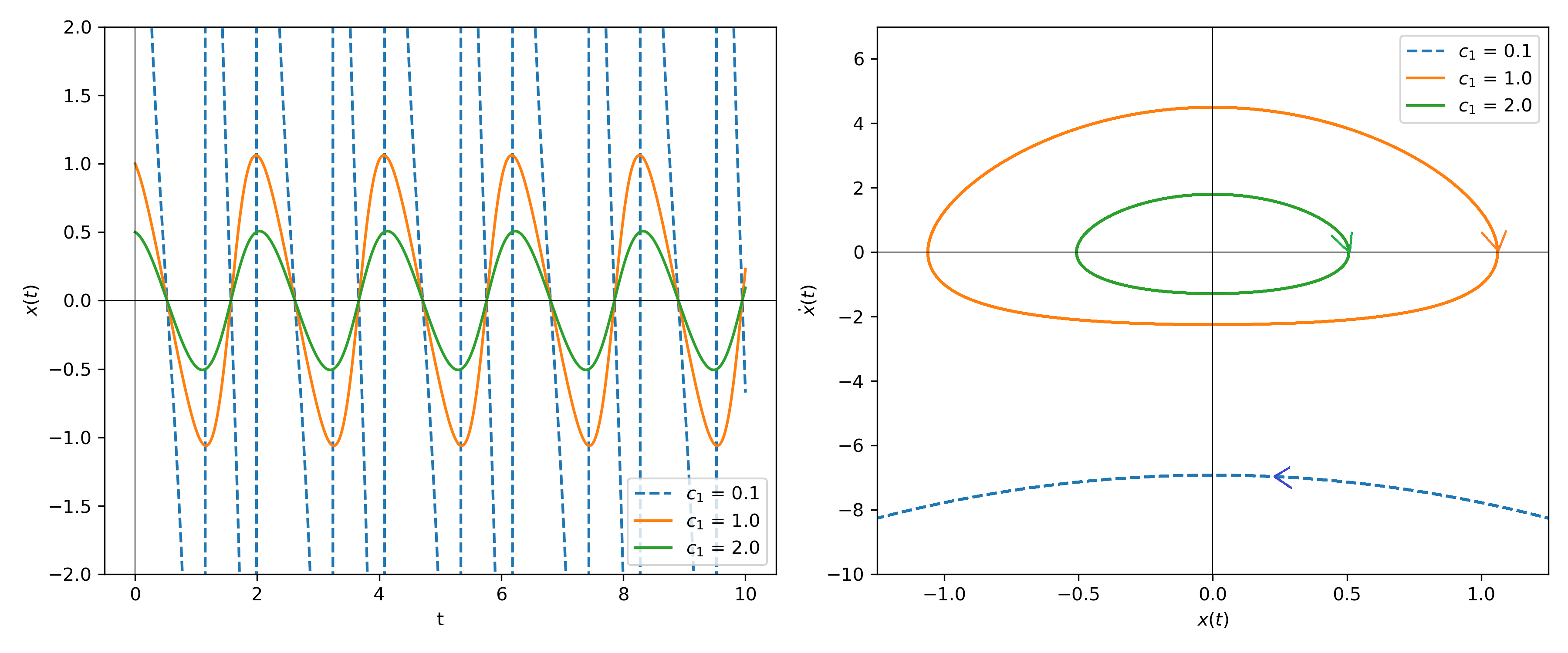}
		\caption{Case $q=1$ for  $k=1,~\omega=3$ and the indicated values of $c_1$. Solutions (\ref{3-18b}) and their phase portraits.}\label{f1}
	\end{figure}

\begin{figure}[H]\centering
        \includegraphics[width=0.85\linewidth]{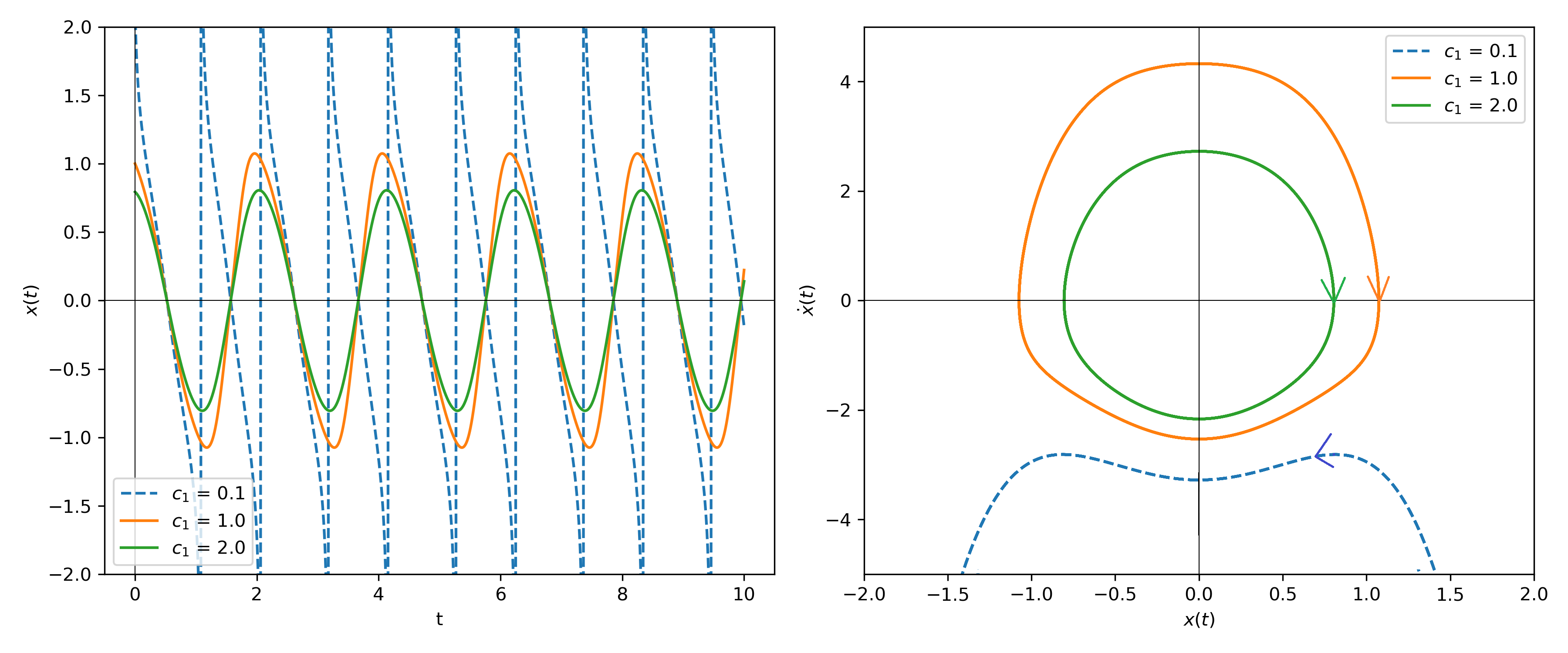}
		\caption{Same as in the previous figure, but for the case $q=3$. 
        }\label{f3}
	\end{figure}

	For $q=1$, we obtain the cubic modified Emden oscillator with linear friction coefficient and associated Bernoulli equation of quadratic power, \cite{pla2025} 
   \begin{equation}\label{3-18}
   \begin{cases}
		&\ddot{x}+3k x \dot{x}+k^2 x^{3}+\omega^2 x=0~,\\
        &\dot{x}+\omega\tan(\omega t)x=-kx^2~,
        \end{cases}
	\end{equation}
whereas for $q=3$, the pair of modified Emden-Bernoulli oscillators is 
	\begin{equation}\label{3-22}
   \begin{cases}
		&\ddot{x}+5k x^3 \dot{x}+k^2 x^{7}+\omega^2 x=0~,\\
        &\dot{x}+\omega\tan(\omega t)x=-kx^4~.
        \end{cases}
	\end{equation}
The solutions are
	\begin{equation}\label{3-18b}
		x(t)=\frac{\cos(\omega t)}{c_1+\frac{k}{\omega}\sin(\omega t)}=\frac{\cos(\omega t)}{c_1+kt\,{\rm sinc}(\omega t)}~
	\end{equation}
and 
	\begin{equation}\label{3-24}
		x(t)=\frac{\cos (\omega t)}{[c_1+kt\left({\frac{9}{4}\,\rm sinc}(\omega t)+\frac{3}{4}{\rm sinc} (3\omega t)\right)]^{1/3}}~, 
	\end{equation}
respectively.
	Solution \eqref{3-18b} is regular and periodic with $T=2\pi/\omega$ if $|c_1|>k/\omega$; singular otherwise.
Solution \eqref{3-24} is also regular and isochronic of the same period if $|c_1|>2k/\omega$, and singular otherwise, and so forth.
Thus a feature of these oscillators, as deduced from \eqref{eq3.44b}, is that the forbidden range of $c_1$ for bounded isochronous solutions grows with the order of nonlinearity, asymptotically as $\sqrt{\pi q/2}\,k/\omega$.
	In figures \ref{f1} and \ref{f3}, one can see plots of these solutions for three different initial conditions: one in the singular regime and two in the bounded isochronous range, along with their phase portraits. The denominator of these solutions presents only oscillatory odd harmonic terms up to a constant determined by the initial condition, and in the phase plane the dynamics is related to periodic motion with closed orbits around a center, or open trajectories corresponding to singular solutions.

\medskip

{\bf The even cases $q=2,4$}\\
	
	For $q=2$, the modified Emden oscillator is a 5th-power oscillator with quadratic friction coefficient and the Bernoulli equation is of cubic power,
	\begin{equation}\label{3-19}
   \begin{cases}
		&\ddot{x}+4k x^2 \dot{x}+k^2 x^{5}+\omega^2 x=0~,\\
        &\dot{x}+\omega\tan(\omega t)x=-kx^3
        \end{cases}
	\end{equation}
while for $q=4$,  one deals with a modified Emden oscillator of 9th power and a corresponding Bernoulli equation of fifth power,  
\begin{equation}\label{3-25}
   \begin{cases}
		&\ddot{x}+6 k x^4 \dot{x}+	k^2 x^9+\omega^2 x=0~,\\
        &\dot{x}+\omega\tan(\omega t)x=-kx^5~.
        \end{cases}
	\end{equation}
	Solutions are 
	\begin{equation}\label{3-21}
		x(t)=\frac{\cos(\omega t)}{[c_1+kt(1+{\rm sinc}(2\omega t))]^{1/2}}~
	\end{equation}
and 
	\begin{equation}\label{3-26}
		x(t)=\frac{\cos (\omega t)}{[c_1+kt \left(\frac{3}{2}+2{\rm sinc} (2 \omega t)+\frac{1}{2}{\rm sinc} (4 \omega t)\right)]^{1/4}}~,
	\end{equation}
respectively.

\begin{figure}[H]\centering
        \includegraphics[width=0.85\linewidth]{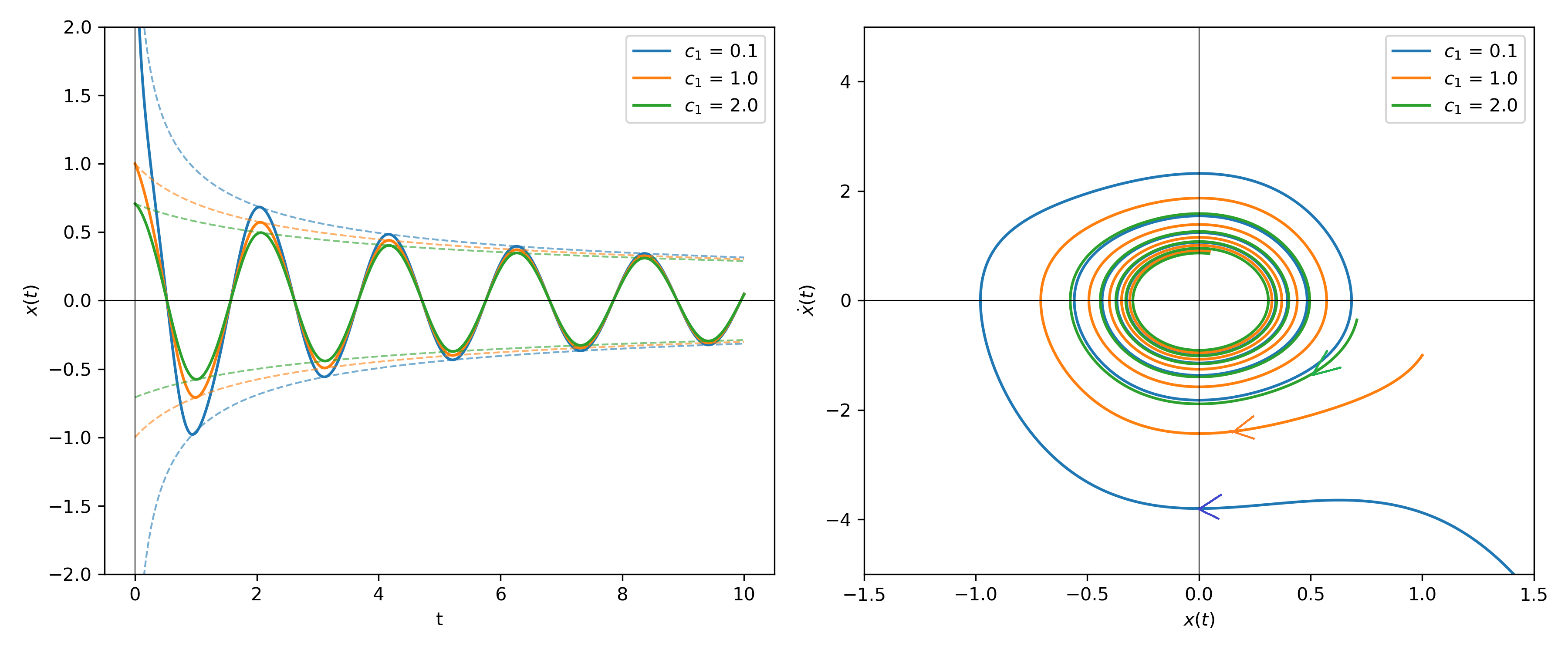}
		\caption{Case $q=2$ for the same numerics as previously. 
        }\label{f2}
	\end{figure}

\begin{figure}[H]\centering
         \includegraphics[width=0.85\linewidth]{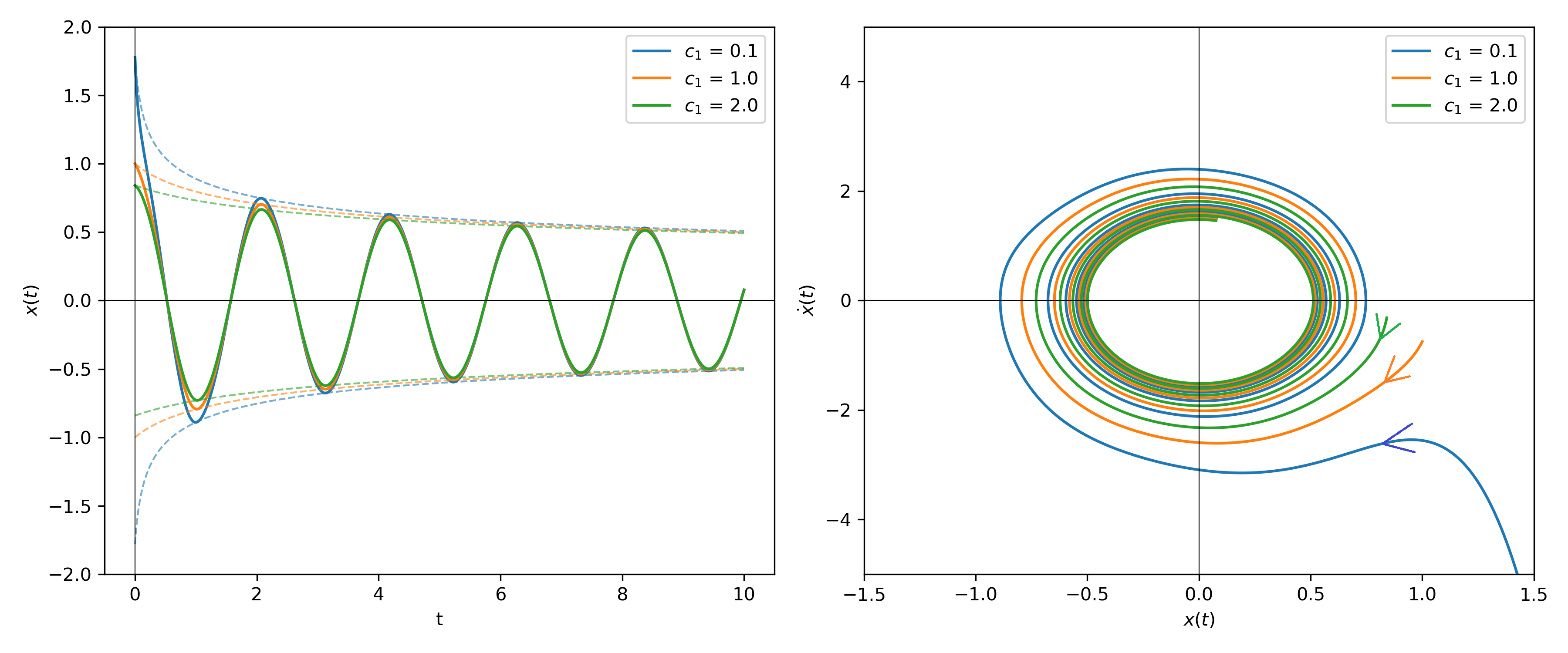}	
		\caption{Case $q=4$ for the same numerics as previously. 
       }\label{f4}
	\end{figure}

	The absolute value of the denominator in these analytic solutions goes to infinity as $t$ increases because of the secular term. This implies that these oscillators are damped ones.
	Plots for these even cases are presented in figures~\ref{f2} and \ref{f4}, respectively, displaying an algebraically damped oscillatory behavior whose envelopes are $E_2=1/\sqrt{kt+c_1}$ and 
$E_4=\bigl(\tfrac{3}{2}kt+c_1\bigr)^{-1/4}$, respectively.
The phase portrait describes a spiral orbit that goes to zero for large values of $t$.

	To end this section, we mention that the damped waveform solutions in the even cases have singularities located at the zeros of their denominators, i.e., as determined
	from the equations $1+{\rm sinc}(2\omega t)=-c_1/kt$ and $3/2+2{\rm sinc} (2 \omega t)+(1/2){\rm sinc} (4 \omega t)=-c_1/kt$ for $q=2$ and $q=4$, respectively.
	The location of the singularities can be found only graphically since the equations are transcendental. In figures~\ref{f2} and \ref{f4}, the singularities do not appear because they are	located on the negative semiaxis. 
	

\numberwithin{equation}{section}
\renewcommand\thesection{\arabic{section}}
\section{Lagrangian formulation}\label{sec5} 
Because of the velocity dependence in the nonlinear equation \eqref{eq0}, a non-standard formulation of the dynamics is required; for this reason we use Lurie's Lagrangian dissipation description \cite{Lurie,Razavy}.
We found that the equation of motion \eqref{eq0} can be derived from a dissipative Lagrangian in Lurie's form: this ODE possesses a single degree of freedom, and the dissipative force is identified with the middle term of \eqref{eq1},
\begin{equation}\label{fd}
	f_{1}(x,\dot{x}) = -(q+2)k\,x^{q}\,\dot{x}~,
\end{equation}
where the subscript indicates the single degree of freedom. Comparing with Lurie's multiple-degree-of-freedom form,
$f_{j} = -k_{j}(x_{1},\cdots,x_{N})\,g_{j}(\dot{x}_{i})$, we identify:
\begin{equation}\label{k}
	k_{1}(x) = (q+2)k\,x^{q}, \qquad g_{1}(\dot{x}) = \dot{x} .
\end{equation}
Since $\dot{x}\,g_{1}(\dot{x}) = \dot{x}^{2} \ge 0$, Lurie's condition is satisfied.
Substituting \eqref{k} into the definition of Lurie's dissipation function \cite{Lurie}, we obtain
\begin{equation}\label{FL}
	\mathcal{F}^{L} = k_{1}(x)\int_{0}^{\dot{x}} y\,dy
	= \frac{1}{2}(q+2)\,k\,x^{q}\,\dot{x}^{2}.
\end{equation}
Thus, the Lagrangian is
$L = \tfrac{1}{2}\dot{x}^{2} - V(x)$, with the potential
\begin{equation}\label{3.20}
	V(x) = \frac{k^{2}x^{2q+2}}{2q+2} + \frac{\omega^{2}x^{2}}{2}~.
\end{equation}
Then the Euler-Lagrange equation with dissipation yields
\begin{equation}\label{3.21}
	\frac{d}{dt}\left(\frac{\partial{L}}{\partial\dot{x}}\right)-\frac{\partial L}{\partial x}+\frac{\partial \mathcal{F}^L}{\partial \dot{x}}=\ddot{x} + k^{2}x^{2q+1} + \omega^{2}x
	+(q+2)k\,x^{q}\dot{x} = 0~,
\end{equation}
which is exactly \eqref{eq0}. \\
Lurie's construction requires $k_{j} > 0$, and here $k_{1}(x) = (q+2)k\,x^{q}$ is positive semidefinite (vanishing only at $x=0$) only for even $q$ and $k>0$.
For odd $q$, the damping term changes sign with $x$: the energy is pumped into the system on half of the orbit, and $\mathcal{F}^{L}$ can no longer be
interpreted as a pure dissipation function, even though \eqref{FL} still reproduces the equation of motion correctly.

It is worth noting that, for constant $k_1$, $\mathcal{F}^L$ reduces to the classical Rayleigh dissipation function \cite{Razavy}; here it is, more precisely, a position-dependent generalization of Rayleigh's quadratic-in-velocity form, with the drag coefficient promoted from a constant to $k_1(x)=(q+2)kx^q$. This is the same construction used, under the name ``Rayleigh force field,'' by Mustafa \cite{Mustafa2023} for the $q=1$ case: his dissipation function $\mathcal{R}(u,\dot u)=\tfrac12\alpha u\dot u^2$, entering the Euler-Lagrange equation exactly as in \eqref{3.21}, coincides with the more general function $\mathcal{F}^L$ for modified Emden oscillators once $\alpha=3k$ is identified as the case $q=1$. 
As Mustafa \cite{Mustafa2023} points out, this standard-Lagrangian route keeps the usual gain-loss correlation between kinetic and potential energy intact ($L=T-V$), in contrast with non-standard constructions in which that correlation is given up.

This choice --- keeping the standard kinetic-minus-potential form $L=\tfrac12\dot x^2-V(x)$ and moving the $x^q\dot x$ term into an external dissipation function --- contrasts with the approach of Bagchi {\em et al.} \cite{bagchi2025}, who treat the $q=1$ case by folding the same term into a non-standard kinetic structure via the Jacobi last multiplier. Their Lagrangian, valid only where the associated Abel equation satisfies the Chiellini integrability condition, is a fractional power of a shifted velocity with no separable $T-V$ split; the resulting Legendre transform yields a fully conservative but branched, momentum-dependent-mass bi-Hamiltonian pair, with no dissipation function at all. That construction is tied to the Chiellini condition and therefore reaches only $q=1$ and a few other isolated values, whereas the Lurie/Rayleigh route used here applies for arbitrary $q$ and is precisely what exposes the odd-even dichotomy: $k_1(x)$ is sign-definite only for even $q$, so only there does $\mathcal{F}^L$ remain a true dissipation function.
\numberwithin{equation}{section}
\renewcommand\thesection{\arabic{section}}
\section{Bendixson-Dulac criterion}\label{sec6}   
Equation~\eqref{eq0} can be written as a first-order system by introducing
$y_1 = x$ and $y_2 = \dot{x}$ (these variables are used only in this section):
\begin{equation}
	\mathcal{L}:\quad
	\left\{
	\begin{aligned}
		\dot{y}_1 &= y_2\,, \\
		\dot{y}_2 &= -\bigl(k(q+2)\,y_1^q\,y_2 + k^2 y_1^{2q+1} + \omega^2 y_1\bigr)\,.
	\end{aligned}
	\right.
\end{equation}
Setting $\dot{y}_2 = 0$ and $\dot{y}_1 = 0$ simultaneously yields $y_2 = 0$ and
\begin{equation}
	x\bigl(k^2 x^{2q} + \omega^2\bigr) = 0\,.
\end{equation}
The dynamical system associated with equation~\eqref{eq0} therefore has a
unique equilibrium at $(x, \dot{x}) = (0, 0)$, and the nonlinear terms determine
the global behavior.

In order to investigate the existence of periodic solutions, we apply the
Bendixson-Dulac criterion \cite{BDC}, which states that if the divergence of
$(\dot{y}_1, \dot{y}_2)$ does not vanish identically and does not change sign in a
simply connected region, then no periodic orbit can lie entirely within that region.
In more general terms, one can use a Dulac function $B(y_1,y_2)$,
i.e.\ a $C^1$ function such that $\nabla\cdot(B\mathcal L)$ does not change sign and does not vanish identically in a simply connected region.
Using $B=1$ (Bendixson's criterion) for the present system the divergence is
\begin{equation}
	\nabla\cdot\mathcal{L} = -k(q+2)\,y_1^{q}\,.
\end{equation}
For the even case $q = 2n$, taking the whole phase plane as the region, the Bendixson's criterion gives
\begin{equation}
	\nabla\cdot\mathcal{L}_{2n} = -2k(n+1)\,y_1^{2n}\,,
\end{equation}
which preserves the same sign for all $y_1 \neq 0$. Therefore, the criterion
excludes the existence of periodic orbits for even $q$. This result is consistent with
the asymptotically damped behavior exhibited by the explicit solutions derived above.
For odd $q = 2n+1$, the divergence
$\nabla\cdot\mathcal{L} = -k(2n+3)\,y_1^{2n+1}$ changes sign on $\mathbb{R}$, so the
criterion is not applicable, and a different approach is used to study the behavior in this case.
\numberwithin{equation}{section}
\renewcommand\thesection{\arabic{section}}
\section{Polar representation}\label{sec7}

A simplified polar representation of these oscillators with $k = \omega = 1$ is due to
Iacono and Russo \cite{IR2011}. For the general case, we introduce $y=\dot x+kx^{q+1}$ that turns \eqref{eq0} into the first-order system
\begin{equation}
	\left\{
	\begin{aligned}
		\dot{x} &= y - k x^{q+1}\,, \\
		\dot{y} &= -\omega^2 x - k x^q y\,.
	\end{aligned}
	\right.
	\label{sys2}
\end{equation}

    Solving $\dot x=\dot y=0$ for $\omega\neq 0$, we found that the origin is the unique real fixed point for every $q$, and the global behaviour is decided by the nonlinear terms.

	For this system we use the elliptic coordinates, which in the case $\omega=1$ recovers the standard polar map.
	\begin{equation}\label{eq:rhodef}
		\rho=\sqrt{\omega^{2}x^{2}+y^{2}}~,\qquad
		x=\frac{\rho}{\omega}\cos\theta~,\qquad y=\rho\sin\theta~,
	\end{equation}
	so that $\tan(\theta)=\frac{y}{\omega x}$. 
	
	From \eqref{sys2} we obtain the relation:
	\begin{equation}\label{eq:cross}
		x\dot y-y\dot=-\left(\omega^{2}x^{2}+y^{2}\right)=-\rho^{2}~,
	\end{equation}
	and the elliptic radial derivative
	\begin{equation}\label{eq:rad}
		\rho\dot{\rho}=\frac{1}{2}\frac{d}{dt}\rho^{2}
		=\omega^{2}x\dot x+y\dot y
		=-k\,x^{q}\bigl(\omega^{2}x^{2}+y^{2}\bigr)
		=-k\,x^{q}\rho^{2}~.
	\end{equation}
	Taking this relation and \eqref{eq:rhodef} the velocity is:
	\begin{equation}\label{dotrho}
		\dot{\rho}=-\frac{k}{\omega^q}\rho^{q+1}\cos^q(\theta)~.
	\end{equation}
	Since $\dot\theta=\omega(x\dot y-y\dot x)/\rho^{2}$, equation \eqref{eq:cross} gives the angular velocity equation
	\begin{equation}\label{eq:polar}
	\dot\theta=-\omega~,	
	\end{equation}
	which is valid for every $q$ and every $k$.

	\subsection{The odd--even dichotomy from the elliptic polar representation}
	
	 From \eqref{eq:polar}, $\theta(t)=\theta_{0}-\omega t$
	for all $q$, $k$ and all initial data: the orbits rotate clockwise at the constant rate $\omega$.
	
	 Since $\dot\theta=-\omega$ is constant, any closed orbit is traversed in
	\begin{equation}
		T=\int_{0}^{2\pi}\frac{d\theta}{|\dot\theta|}=\frac{2\pi}{\omega}~,
	\end{equation}
	independently of the amplitude, $c_{1}$, and $k$. This is the isochronicity established analytically in \eqref{odd}.
	
\begin{figure}[H]\centering
		\includegraphics[width=0.65\linewidth]{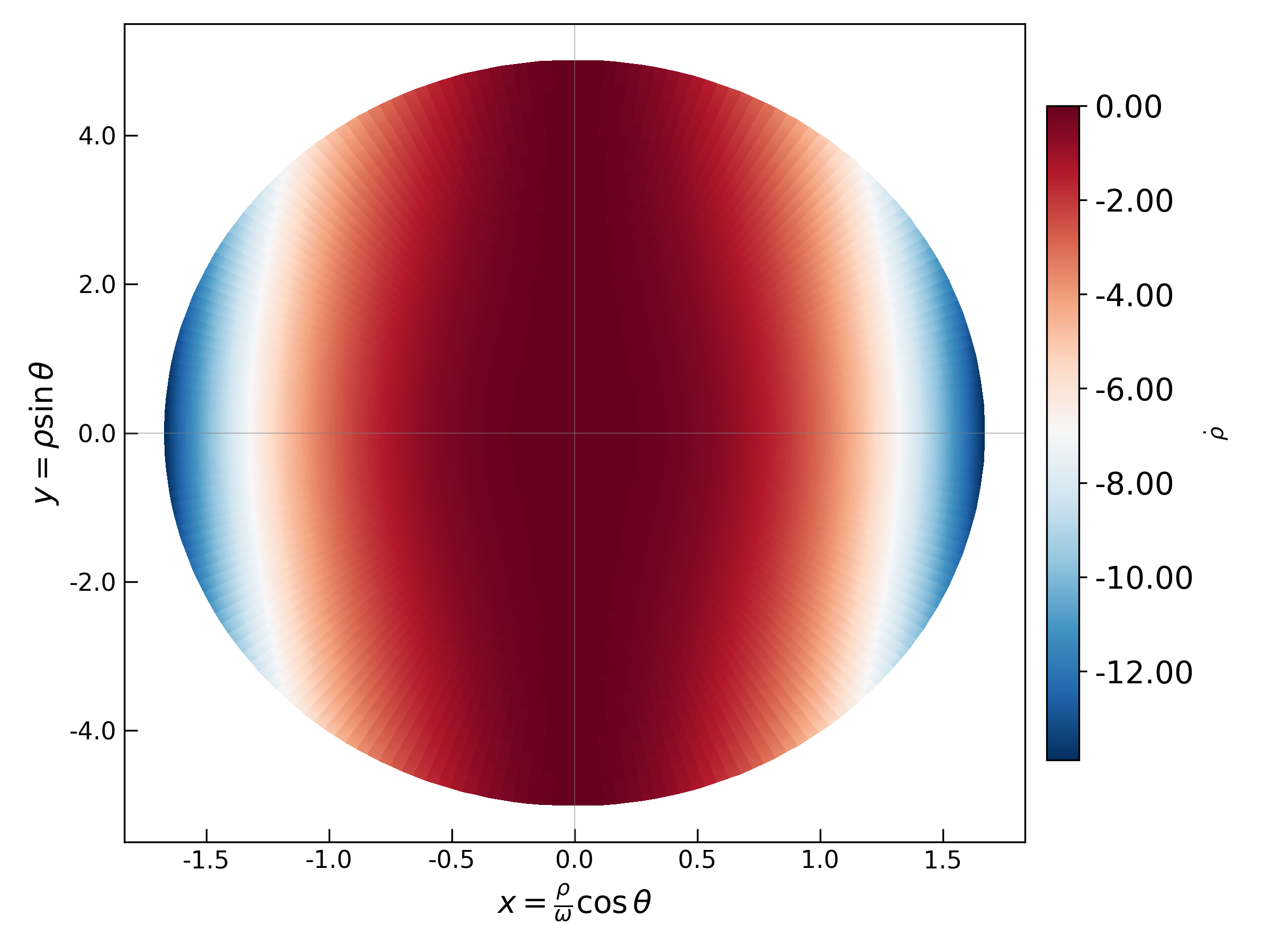}
			\caption{Sign of the radial velocity $\dot\rho=-(k/\omega^{q})\rho^{\,q+1}\cos^{q}\theta$         
			in the $((\rho/\omega)\cos\theta,\rho\sin\theta)$ plane, for $k=1$, $q=2$, $\omega=3$, $\rho\in[0,5]$,
			$\theta\in[0,2\pi]$. }
		\label{fp2}
	\end{figure}

    For even $q$ and $k>0$ we have that $x^{q}\ge 0$, so
	\eqref{dotrho} shows that the radial velocity $\dot{\rho}$ is non-positive for all $t$ (vanishing only at isolated instants where $\theta=\pi/2,3\pi/2$), so that $\rho(t+\Delta t)\le \rho(t)$ for any $t$; hence, for large values of the evolution parameter, the trajectories converge to the fixed point at the origin, indicating damped oscillation. This behavior is shown for the case $q=2$ in the figure \ref{fp2}, where the density plot $(x,y)$ map shows negative radial velocities for all values of $\theta$ and $\rho$. 
	
	 For $q=2n+1$ the factor $\cos^{q}\theta$ in \eqref{dotrho} changes sign twice per revolution: $\rho$ decreases on the half-orbit
	where $x>0$ and increases on the half-orbit where $x<0$. This is consistent with
Section \ref{sec5}, where $k_{1}(x)=(q+2)kx^{q}$ fails to be
	positive semidefinite for odd $q$, so that energy is pumped into the system on half of the
	orbit and $\mathcal{F}^{L}$ loses its interpretation as a dissipation function.
	
In this case the radial velocity changes sign depending on $\theta$, which implies
that the distance from a given point on the orbit to the fixed point may increase for
some values of $\theta$ and decrease for others.

\begin{figure}[H]\centering
		\includegraphics[width=0.65\linewidth]{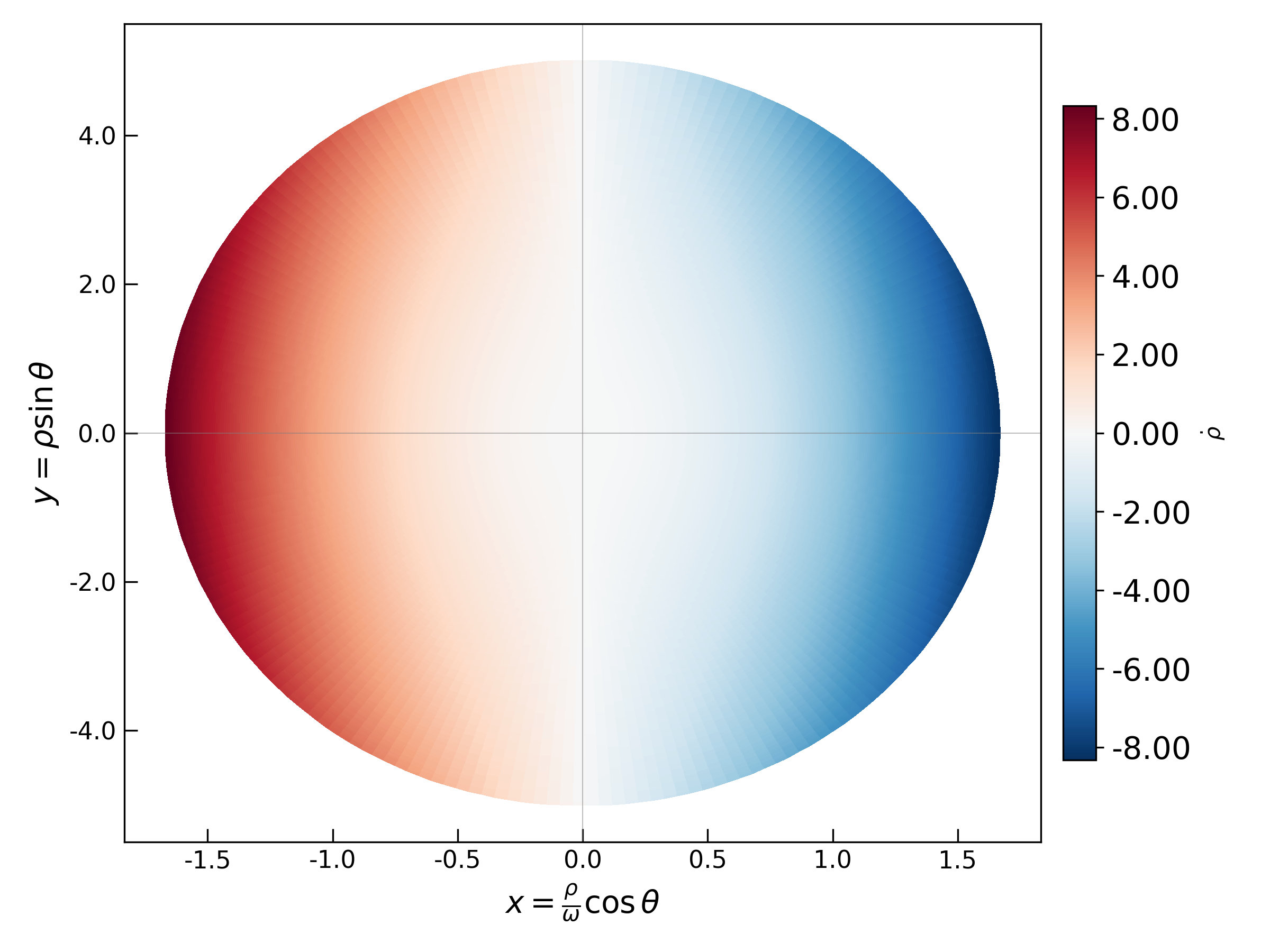}
		\caption{Sign of the radial velocity $\dot\rho=-(k/\omega^{q})\rho^{q+1}\cos^{q}\theta$ in the $((\rho/\omega)\cos\theta,\rho\sin\theta)$ plane, for $k=q=1$, $\omega=3$, $\rho\in[0,5]$, $\theta\in[0,2\pi]$. }
		\label{fp1}
	\end{figure}

 This behavior can be analyzed in a density plot, as in Fig.~\ref{fp1}, which shows the $\dot{\rho}$ value for $\theta\in[0,2\pi]$ and $\rho\in[0,5]$; this plot shows that the sign of the radial velocity changes throughout the evolution of $\theta$ and $\rho$.
 
\subsection{Extension to the $c$-deformed equation}\label{sec:polarc}

The same construction applies to \eqref{eq:underdamped_lienard}. With $Y=\dot x+kx^{q+1}+cx$, $\rho_c^2=\omega^2x^2+Y^2$ and $\tan\theta_c=Y/(\omega x)$, one finds, for all $q$, $k$ and $c$,
\begin{equation}\label{eq:polarc}
\dot\theta_c=-\omega~,\qquad \dot\rho_c=-c\,\rho_c-\frac{k}{\omega^q}\,\rho_c^{\,q+1}\cos^q\theta_c~.
\end{equation}
The phase therefore rotates uniformly also in the damped case; equivalently,
\begin{equation}\label{eq:arctanI}
\arctan\frac{\dot x/x+kx^q+c}{\omega}+\omega t={\rm const}
\end{equation}
is a time-dependent first integral for every $q$. The radial equation in \eqref{eq:polarc} is of Bernoulli type, with solution
\begin{equation}\label{eq:rhoc}
\rho_c^{-q}=e^{qct}\Big[C+\frac{qk}{\omega^q}\int e^{-qct}\cos^q(\theta_0-\omega t)\,dt\Big]~,
\end{equation}
equivalent to the general solution \eqref{eq:gensol} of Section \ref{sec9}.
\numberwithin{equation}{section}
\renewcommand\thesection{\arabic{section}}
\section{First integrals, Jacobi last multipliers and Lagrangians} \label{sec8} 

In this section, we follow closely \cite{Sin2020} to use linearization through Sundman transformations  
as a systematic way to obtain first integrals of those Li\'enard equations that are linearizable by such maps. 

\medskip 


\subsection{The case $q=1$: Sundman linearization, first integral and Lagrangian}

In \cite{Sin2020}, the Li\'enard equation \eqref{eq1} and the autonomous Sundman transformations are written as
	\begin{equation}\label{eq:s1}
		y_{zz} +g(y)y_z + h(y) = 0~,
	\end{equation}
and
	\begin{equation}\label{eq:s2}
		w={\cal F}(y)~, \qquad d\xi ={\cal G}(y)dz~,
	\end{equation}
respectively. The Sundman functions ${\cal F}$ and ${\cal G}$, supposed sufficiently smooth, satisfy ${\cal F}_y{\cal G}\neq 0$. The condition
	\begin{equation}\label{eq:condition}
{\cal G}{\cal F}_{yy}-{\cal F}_y{\cal G}_y=0~,
\end{equation}
which removes the $y_z^2$ term, is necessary for the standard Li\'enard equation \eqref{eq:s1} to be mapped onto the linear equation
	\begin{equation}\label{eq:s3}
		w_{\xi\xi} +\tilde{\beta} w_\xi+\alpha w= 0~,
	\end{equation}
where $\alpha\neq 0$ and $\tilde{\beta}\neq 0$ are arbitrary real parameters; the coefficients $g$ and $h$ must then have the form \eqref{eq:Sinh} below. The linear damped oscillator \eqref{eq:s3} has the autonomous first integral
	\begin{equation}\label{eq:s4}
		I_w=(2w_\xi+(\tilde{\beta}+\tilde{\rho}) w)^{\tilde{\beta}+\tilde{\rho}}/(2w_\xi+(\tilde{\beta}-\tilde{\rho}) w)^{\tilde{\beta}-\tilde{\rho}}~,
	\end{equation}
with $\tilde\rho=\sqrt{\tilde\beta^2-4\alpha}$, which helps to obtain first integrals of Li\'enard equations. Since its exponents are in general irrational, $I_w$ is transcendental; it becomes rational when $\tilde\rho/\tilde\beta$ is rational (Corollary 2.3 of \cite{Sin2020}).

Condition \eqref{eq:condition} is equivalent to $\partial_y({\cal F}_y/{\cal G})=0$, so that ${\cal G}=A{\cal F}_y$, with $A$ a nonzero constant since ${\cal F}$ and ${\cal G}$ depend on $y$ only. Substitution yields
$y_{zz}+\tilde{\beta} A{\cal F}_y\,y_z+\alpha A^2{\cal F}{\cal F}_y=0$. Hence $g=\tilde{\beta} A{\cal F}_y$, $\frak{g}+\mu=\tilde{\beta} A{\cal F}$, where $\frak{g}$ is a primitive of $g$, $\frak{g}_y=g$, and
\begin{equation}\label{eq:Sinh} 
h=\frac{\alpha}{\tilde{\beta}^2}\,(\frak{g}+\mu)\,g . 
\end{equation}
Eliminating $\mu$ gives the condition 
\begin{equation}\label{eq:coroll2.2}   
\tilde{\beta}^2(hg_y-gh_y)+\alpha g^3=0~ 
\end{equation}
for Li\'enard equations to have the first integral   
\begin{equation}\label{eq:Sincor} 
I_L=\big(2\tilde{\beta} y_z+(\tilde{\beta}+\tilde{\rho}_L)(\frak{g}+\mu)\big)^{\tilde{\rho}_L+\tilde{\beta}}\big(2\tilde{\beta} y_z+(\tilde{\beta}-\tilde{\rho}_L)(\frak{g}+\mu)\big)^{\tilde{\rho}_L-\tilde{\beta}},\qquad \tilde{\rho}_L=\sqrt{\tilde{\beta}^2-4\alpha}~. 
\end{equation}

For $g=3kx$, $h=k^2x^3+\lambda x$, one finds $hg_x-gh_x=-6k^3x^3$, independently of $\lambda$. The condition holds if and only if
\begin{equation}\label{8.9}
\alpha=\tfrac29\tilde{\beta}^2,\qquad\text{for all }k\neq0\text{ and all }\lambda,
\end{equation}
so that $\tilde{\rho}_L=\tilde{\beta}/3$, and \eqref{eq:Sinh} fixes $\mu=3\lambda/(2k)$. With $\tilde{\beta}=3$, $\alpha=2$:
\begin{equation}
w=k^2x^2+\lambda,\qquad \dd\xi=kx\,\dd t\qquad\Longrightarrow\qquad w_{\xi\xi}+3w_\xi+2w=0 .
\end{equation}
Indeed $w_\xi=2k\dot x$ and $w_{\xi\xi}=2\ddot x/x$, so multiplying the linear equation by $x/2$ returns the $\lambda$-modified Emden equation. Because $\tilde{\rho}_L/\tilde{\beta}=1/3$ is rational, Corollary 2.3 of \cite{Sin2020} gives a rational integral. For $w=ae^{-\xi}+be^{-2\xi}$ the ratio $(w_\xi+2w)^2/(w_\xi+w)=-a^2/b$ is constant, hence
\begin{equation}
I_E=\frac{(k\dot x+k^2x^2+\lambda)^2}{2k\dot x+k^2x^2+\lambda}
\end{equation}
is an autonomous rational first integral for every $\lambda$.

\medskip

A Lagrangian exists for any second-order ordinary differential equation \cite{Bolza1904}, and it can be obtained explicitly from a Jacobi last multiplier $M$ through $M=\partial^2L/\partial\dot x^2$, i.e. by two quadratures \cite{Nucci2021}.
For a planar autonomous system, a first integral $I(x,\dot x)$ yields the Jacobi last multiplier (JLM) $M=\dot x^{-1}\,\partial I/\partial\dot x$, unique up to a factor $\Psi(I)$. Write now $I_E=P^2/Q$ with $P=k\dot x+k^2x^2+\lambda$ and $Q=2k\dot x+k^2x^2+\lambda$. Since $Q-P=k\dot x$,
\[
\frac{\partial I_E}{\partial\dot x}=\frac{2kP(Q-P)}{Q^2}=\frac{2k^2\dot x\,P}{Q^2}~,
\qquad\text{so}\qquad
M=\frac{k\dot x+k^2x^2+\lambda}{(2k\dot x+k^2x^2+\lambda)^2}
\]
is a JLM up to a constant. 
Also, through the Darboux factor method, along solutions of the $\lambda$-modified Emden equation,
\begin{equation}\label{eq:dotP}
\dot P=k\ddot x+2k^2x\dot x=-kx\,P ,
\end{equation}
so $P$ is a Darboux function with cofactor $-kx$. Writing \eqref{eq01} as $\ddot x={\cal N}(x,\dot x)$, a JLM satisfies $\frac{d}{dt}\ln M=-\partial{\cal N}/\partial\dot x=3kx$, so the simpler function
\begin{equation}\label{eq:MP3}
M=(k\dot x+k^2x^2+\lambda)^{-3} 
\end{equation}
is also a JLM. The two multipliers differ by the factor $I_E^2$, a first integral. 

Integrating $\partial^2 L/\partial\dot x^2=M$ twice \cite{Nucci2021} gives
\begin{equation}\label{eq:L}
L=\frac{1}{2k^2\,(k\dot x+k^2x^2+\lambda)}~.
\end{equation}
Indeed $\frac{\dd}{\dd t}L_{\dot x}=(\ddot x+2kx\dot x)/P^3$ and $L_x=-x/P^2$, and multiplying the Euler--Lagrange equation by $P^3$ returns the $\lambda$-modified Emden equation exactly, with no additional potential term. This is essentially the known Lagrangian on page 4 of the 2005 paper of Chandrasekar, Senthilvelan and Lakshmanan~\cite{CSL2005}. It complements the Lurie construction of~section V and the Jacobi-multiplier route of Bagchi \emph{et al.}~\cite{bagchi2025}.

\begin{remark}
The Sundman time $\xi=k\int x\,\dd t$ is exactly the exponent of the CSKL map, $U=x\,e^{\xi}$. The CSKL map sends \eqref{eq01} to $\ddot U+\lambda U=0$, with $\lambda$ in the linear equation, whereas the Sundman map sends it to the fixed overdamped equation $w''+3w'+2w=0$, with $\lambda$ absorbed into $w=k^2x^2+\lambda$. The isochronous oscillations for $\lambda>0$ are invisible in $\xi$: since $\dd\xi=kx\,\dd t$ changes sign whenever $x$ crosses zero, the Sundman map is only local in $t$.
\end{remark}

For the generalized modified Emden equation CSKL (9), $\ddot x+(\frak{c}_1+3kx)\dot x+k^2x^3+\frak{c}_1kx^2+\frak{c}_2x=0$, the coefficients of $x^3$, $x^2$ and $x$ in \eqref{eq:coroll2.2} all give $\alpha=2\tilde{\beta}^2/9$, and the constant term adds $\frak{c}_1(2\frak{c}_1^2/9-\frak{c}_2)=0$. The Sundman route therefore applies if and only if $\frak{c}_1=0$ or $\frak{c}_2=2\frak{c}_1^2/9$. The second case is precisely \eqref{A2} of the Appendix, i.e.\ \eqref{eq01} shifted by $x\to x+\mu$ with $\lambda=-\mu^2/9$. For other values of $\frak{c}_2$, CSKL (9) is CSKL-linearizable but not by an autonomous Sundman map onto a constant-coefficient equation.

\subsection{General $q$: polar first integral, Darboux polynomials and multipliers}

\emph{Sundman maps for general $q$.} For \eqref{eq:underdamped_lienard}, condition \eqref{eq:Sinh} fixes $\alpha/\tilde\beta^2=(q+1)/(q+2)^2$ from the top-order term, and the remaining terms require $\mu=0$ and $\omega^2=-q^2c^2/(q+2)^2$, except for $q=1$, $c=0$, where $\mu=3\omega^2/(2k)$ is allowed. Hence, for $c=0$, the Sundman route covers \eqref{eq0} for all $\omega$ only when $q=1$. For $q\ge2$ it applies only to $\omega=0$, where $w=kx^{q+1}$, $\dd\xi=kx^q\dd t$, $w_{\xi\xi}+(q+2)w_\xi+(q+1)w=0$, with the rational first integral $(\dot x+kx^{q+1})^{q+1}/[(q+1)\dot x+kx^{q+1}]$. The isochronous regime for $q\ge2$ therefore requires other tools, which we now describe.

\emph{Polar first integral.} With the elliptic variables of Section VII, $\dot\theta=-\omega$ and \eqref{dotrho} imply that
\begin{equation}\label{eq:Iq}
I_q=\rho^{-q}+\frac{qk}{\omega^{q+1}}\,S_q(\theta)~,\qquad S_q(\theta)=\int_0^\theta\cos^q s\,ds~,
\end{equation}
is a first integral, since $\frac{d}{dt}\rho^{-q}=\frac{qk}{\omega^q}\cos^q\theta$ and $\frac{d}{dt}S_q=-\omega\cos^q\theta$. For odd $q$, $S_q$ is $2\pi$-periodic and $I_q$ is a global first integral on the punctured plane, whose level curves are the isochronous closed orbits. For even $q$, $S_q$ contains the secular term $\binom{q}{q/2}2^{-q}\theta$, so $I_q$ is multivalued and no global first integral exists, in agreement with the Bendixson--Dulac result of Section \ref{sec6}. 

\emph{Darboux polynomials and Jacobi last multiplier.} The complex conjugate polynomials $\chi_\pm=\dot x+kx^{q+1}\pm i\omega x$ satisfy $\dot\chi_\pm=(-kx^q\pm i\omega)\chi_\pm$, so that
\begin{equation}\label{eq:Darb}
{\cal D}=\chi_+\chi_-=\omega^2x^2+(\dot x+kx^{q+1})^2=\rho^2~,\qquad \dot{\cal D}=-2kx^q\,{\cal D}~,
\end{equation}
is a real Darboux polynomial for every $q$. Since a JLM of \eqref{eq0} must satisfy $\frac{d}{dt}\ln M=(q+2)kx^q$, matching cofactors gives
\begin{equation}\label{eq:Mq}
M_q={\cal D}^{-\frac{q+2}{2}}=\Big[\omega^2x^2+(\dot x+kx^{q+1})^2\Big]^{-\frac{q+2}{2}}~,
\end{equation}
which coincides, up to a constant, with $\dot x^{-1}\partial I_q/\partial\dot x$. The multiplier $M_q$ is single-valued for every $q$, even when no global first integral exists, and $1/M_q$ is an inverse integrating factor, polynomial for even $q$, whose only real zero is the origin. Integrating $\partial^2L/\partial\dot x^2=M_q$ twice gives a Jacobi Lagrangian for every $q$.

\emph{Odd $q$: rational first integral.} For $q=2n+1$, the polynomial
\begin{equation}\label{eq:Nq}
N_q=1+\frac{qk}{\omega^{q+1}}\sum_{j=0}^{n}\frac{(-1)^j}{2j+1}\binom{n}{j}\,(\dot x+kx^{q+1})^{2j+1}\,{\cal D}^{\,n-j}
\end{equation}
equals ${\cal D}^{q/2}I_q$ and satisfies $\dot N_q=-qkx^q\,N_q$. Hence $N_q^2/{\cal D}^{\,q}$ is a rational global first integral and $N_q^{-(q+2)/q}$ is a JLM. For $q=1$ and $\lambda=\omega^2$, $N_1=P/\omega^2$, so $N_1^2/{\cal D}$ is equivalent to $I_E$ and $N_1^{-3}$ reproduces \eqref{eq:MP3}; for $q=3$, $N_3=1+\frac{k}{\omega^4}(\dot x+kx^4)\big(3\omega^2x^2+2(\dot x+kx^4)^2\big)$.

\emph{Damped case.} For \eqref{eq:underdamped_lienard}, with the variables of Section~\ref{sec:polarc}, a JLM must satisfy $\frac{d}{dt}\ln M=2c+(q+2)kx^q$, which is solved by $M_c=e^{-qct}\rho_c^{-(q+2)}$, an autonomous but multivalued function of $(\rho_c,\theta_c)$ through $t=(\theta_0-\theta_c)/\omega$.
\numberwithin{equation}{section}    
\renewcommand\thesection{\arabic{section}}
	\section{The $c$-deformed modified Emden oscillators: damped waveforms and dynamical regimes}\label{sec9}

	We now introduce a real constant $c>0$ into the commutative factorization operators 
	\begin{equation}
		A = D + kx^q + c + i\omega, \qquad A^\dagger = D + kx^q + c - i\omega~. 
	\end{equation}
	In this case, $AA^\dagger x = 0$ generates the Li\'enard equation \eqref{eq:underdamped_lienard} commented at the beginning of the paper, 
with $f(x)=2c+(q+2)kx^q$ and $g(x)=kx^{q+1}(kx^q+2c)+(\omega^2+c^2)x$. 
Setting $c=0$ recovers exactly the equation~\eqref{eq0}, confirming that $c$ is a deformation parameter that continuously connects the two families.
	
		The Jacobian at the origin for arbitrary $c$ is
	\begin{equation}
		J\big|_{(0,0)} = \begin{pmatrix} 0 & 1 \\ -(\omega^2+c^2) & -2c \end{pmatrix},
	\end{equation}
with eigenvalues $\lambda_\pm = -c \pm i\omega$. Since $c>0$, both eigenvalues have strictly negative real part, so the origin is a stable spiral: all trajectories converge to it, 
in contrast to the closed orbits of the isochronous case. For $c>0$ and $k>0$, one can show that the origin is the only fixed point of the system.

	Due to the commutativity relation of the factorization operators $AA^\dagger x = A^\dagger Ax$, we can derive two equivalent expressions for the common factor $e^{-ct}e^{-k\int x^q dt}$. Equating them yields a Bernoulli equation
	\begin{equation}\label{eq:bernoulli_underdamped}
		\dot{x}+\bigl(c-\omega\cot(\omega t)\bigr)x = -kx^{q+1}.
	\end{equation}
	This generalizes the isochronous Bernoulli equation~\eqref{qB}: the tangent is displaced by $\pi/2$ to obtain the minus cotangent and the damping constant $c$ appears explicitly. The substitution $u=x^{-q}$ linearizes~\eqref{eq:bernoulli_underdamped} to
	\begin{equation}
		\dot{u}+q\bigl[\omega\cot(\omega t)-c\bigr]u = kq,
	\end{equation}
	with integrating factor $\sin^q(\omega t)\,e^{-cqt}$. Integration yields the general solution
	\begin{equation}\label{eq:gensol}
		x_q(t) = \frac{\sin(\omega t)}{\left(c_1e^{cqt}+\dfrac{kq}{(2i)^q}\displaystyle\sum_{j=0}^{q}\binom{q}{j}(-1)^j\dfrac{e^{i\omega t(q-2j)}}{i\omega(q-2j)-cq}\right)^{1/q}}~.
	\end{equation}
	The factor $(2i)^q$ combined with the binomial sum converts the complex exponentials into real trigonometric-exponential combinations, as verified in the examples below. The long-time behavior is determined by the dominance of $c_1$ in the denominator as the sum decays exponentially, giving, for $c_1>0$ (which is required for even $q$), an envelope
	\begin{equation}
		E_c(t) \sim \frac{e^{-ct}}{c_1^{1/q}}, \qquad t \to \infty.
	\end{equation}
	The decay rate $c$ does not depend on the nonlinearity order $q$, while only the scaling factor $c_1^{1/q}$ carries a $q$-dependence.

	\subsection{Examples}
	
	We present explicit solutions for $q\in\{1,2,3,4\}$ and compare them with their isochronous counterparts. All underdamped plots use $\omega=4$, $k=c_1=1$ and a set of four numerical values of $c$.

{\bf Case $q=1$}
	
\begin{figure}[H]
		\centering
		\includegraphics[width=0.78\linewidth] {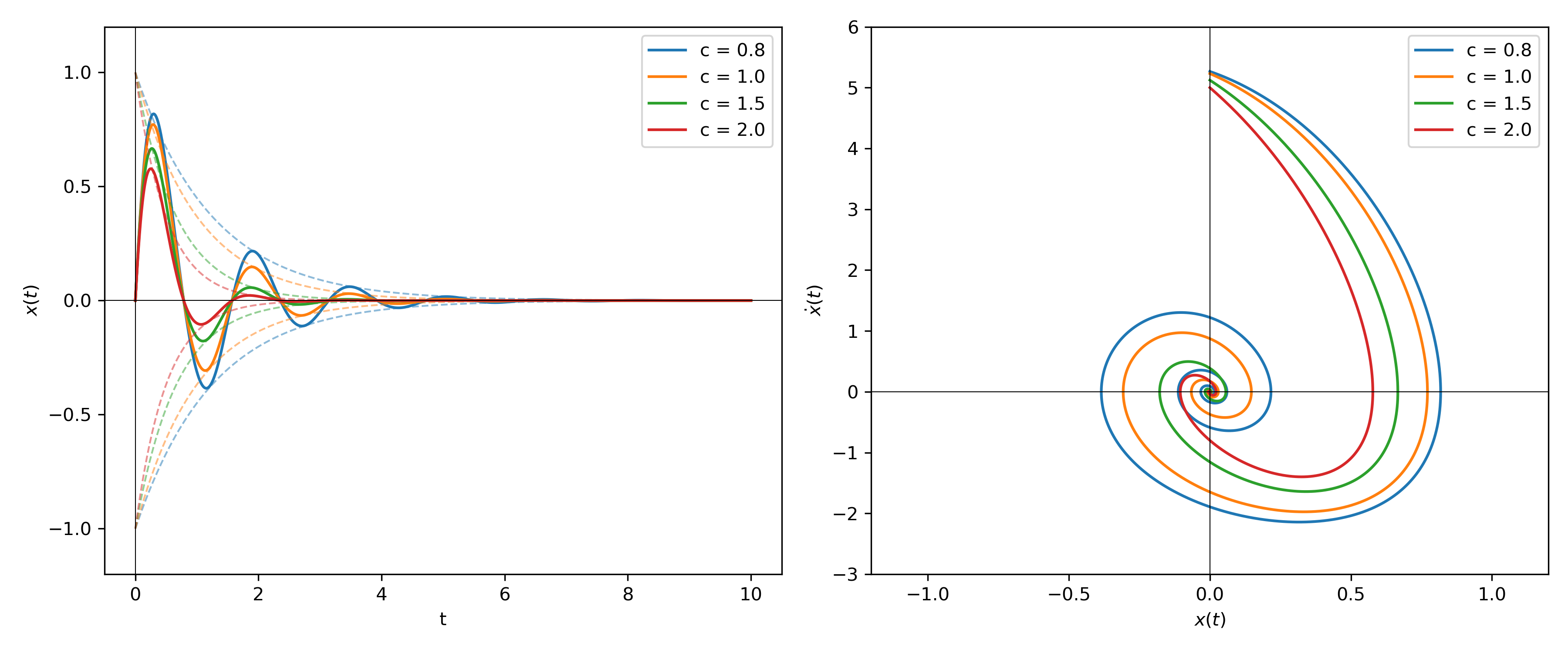}           
		\caption{Case $q=1$ with $\omega=4$, $k=c_1=1$ and $c$ = 0.8, 1.0, 1.5, 2.0. Left: solution $x_1(t)$ showing damped oscillations. Right: phase portrait showing a stable spiral converging to the origin.}
		\label{fig:q1}
	\end{figure}

The corresponding underdamped Li\'enard equation~\eqref{eq:underdamped_lienard} for $q=1$ admits the closed-form solution
	\begin{equation}\label{eq:q1_ud}
		x_1(t;c) = \frac{\sin(\omega t+\delta)}{c_1 e^{ct}-\frac{k}{\omega\sqrt{1+\left(\frac{c}{\omega}\right)^2}}\cos(\omega t+\delta-\phi_1)}~, 
\quad \phi_1={\rm arctan}\left(\frac{c}{\omega}\right)~,
	\end{equation}
	whose amplitude envelope decays as $e^{-ct}$ and whose phase portrait shows a stable spiral converging to the origin. This solution is equivalent to the isochronous case when $\delta=\pi/2$
and $c=0$. Figure~\ref{fig:q1} illustrates the solution and the phase portrait for the chosen parameters. This figure shows that for small values of the parameter $c$ the solution's amplitude near the origin increases and the amplitudes for the next local maxima are bigger than the amplitudes for large values of $c$. This is because in the limiting case $c\to 0$ the local maxima of all the signal converges at the same value, i.e. the periodic regime. 
	
{\bf Case $q=3$}
	
\begin{figure}[H]
		\centering
		\includegraphics[width=0.78\linewidth]{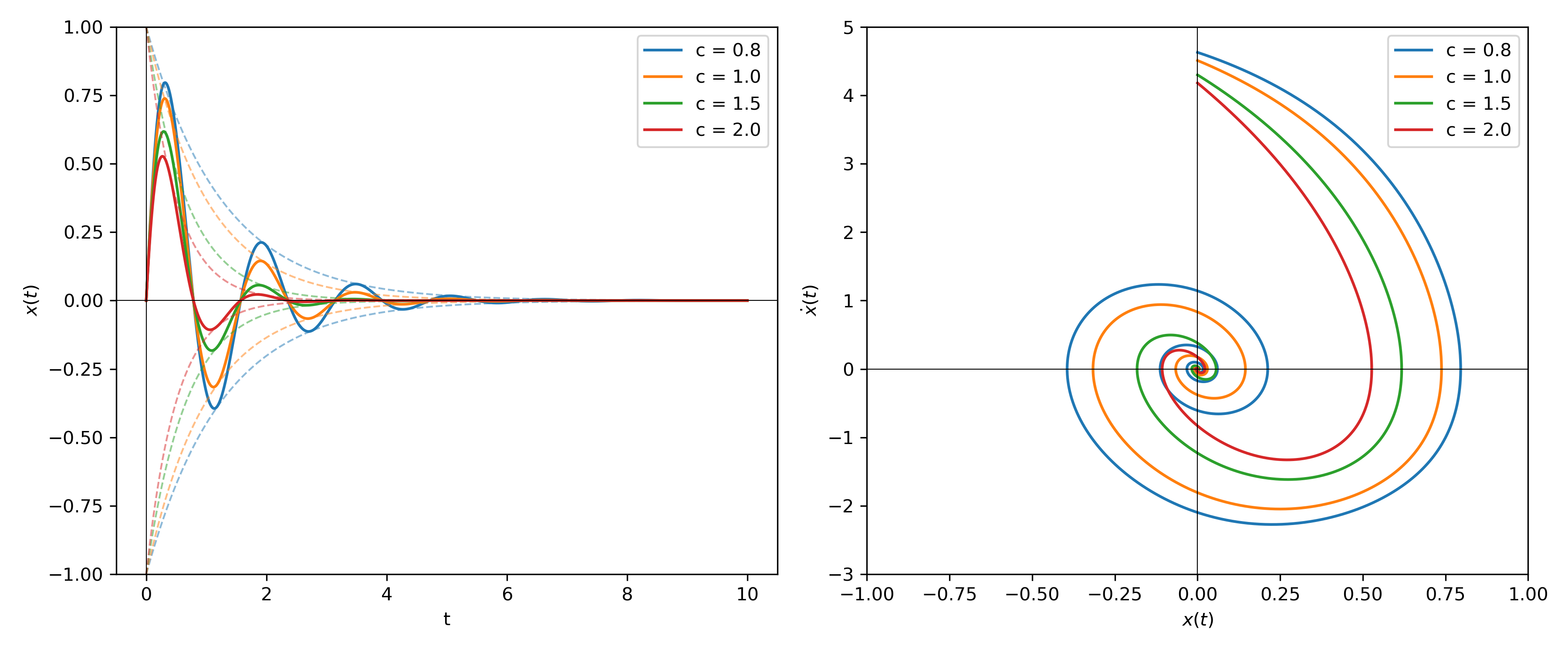}     
		\caption{Case $q=3$ with $\omega=4$, $k=c_1=1$ and and $c$ = 0.8, 1.0, 1.5, 2.0. Left: solution $x_3(t)$ with smooth exponential decay governed by $c$. Right: phase portrait showing convergent spiral to the origin.}
		\label{fig:q3}
	\end{figure}

The underdamped waveform solution of equation~\eqref{eq:underdamped_lienard} with $q=3$, has the closed-form expression
	%
	\begin{equation}\label{eq:q3_ud}
		x_3(t;c) = \frac{\sin(\omega t+\delta)}{\left(c_1e^{3ct} - \dfrac{k}{4\omega}   
\!\left[
\dfrac{3\cos(\omega t+\delta-\phi_3)}{\sqrt{\left(\frac{1}{3}\right)^2+\left(\frac{c}{\omega}\right)^2}}  
-\dfrac{\cos(3(\omega t+\delta)-\phi_1)}{\sqrt{1+\left(\frac{c}{\omega}\right)^2}}\right]\right)^{\frac{1}{3}}}~,  \quad \phi_3={\rm arctan}\left(\frac{3c}{\omega}\right)~,
	\end{equation}
	%
which reduces to the sinc-shape isochronous expression as $c\to 0$ and $\delta=\pi/2$. The cubic root allows negative arguments, so the solution remains real; however, zeros of the denominator are true singularities (see the end of this subsection). Figure~\ref{fig:q3} shows the solution and phase portrait.
	
{\bf Case $q=2$}

	The case $q=2$ has a smooth limit $c\to0$ once the integration constant is redefined as $c_1\to c_1+k/(2c)$, and with $\delta=\pi/2$ it then recovers the algebraically damped solution \eqref{3-21}, or \eqref{even} with $n=1$. 
The underdamped waveform reads:
\begin{equation}\label{eq:q2}
	x_2(t;c) = \frac{\sin(\omega t+\delta)}{\left(c_1e^{2ct}  - \frac{k}{2\omega}  \left[ \frac{\cos(2(\omega t+\delta)-\phi_2)}{\sqrt{1+\left(\frac{c}{\omega}\right)^2}} + \frac{\omega}{c} \right] \right)^{1/2} }~,  \quad \phi_2=\frac{\pi}{2}+{\rm arctan}\left(\frac{c}{\omega}\right)~.
\end{equation}
	
\begin{figure}[H]
	\centering
    \includegraphics[width=0.78\linewidth]{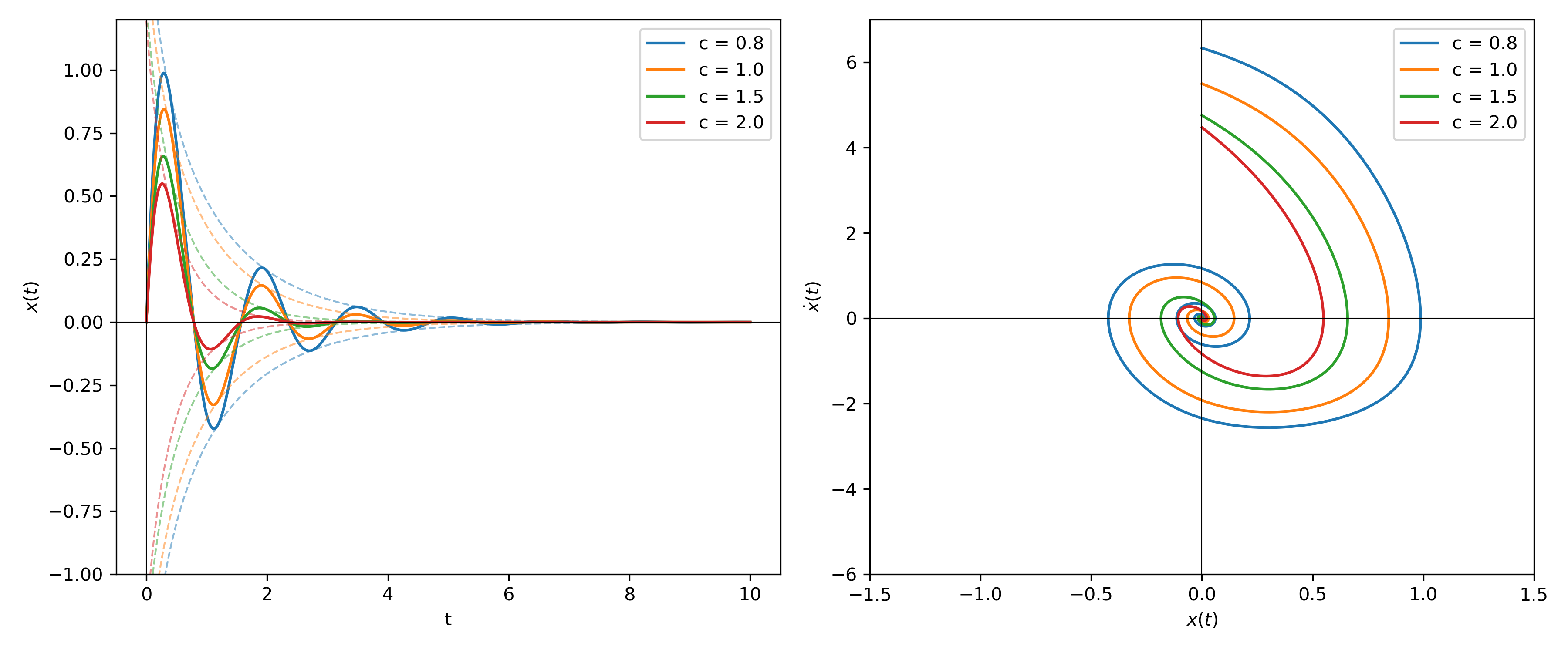}       
	\caption{Case $q=2$ with $\omega=4$, $k=c_1=1$ and $c$ = 0.8, 1.0, 1.5, 2.0. Left: solution $x_2(t)$ with smooth exponential decay governed by $c$. Right: phase portrait showing convergent spiral to the origin.}
	\label{fig:q2}
\end{figure}

	Figure \ref{fig:q2} shows this underdamped waveform and its phase portrait for the chosen parameters.
	
{\bf Case $q=4$}

	For this case, the waveform reads:
\begin{equation}\label{eq:q4}
	x_4(t;c) = \frac{\sin(\omega t+ \delta)}{\left(c_1e^{4ct}-\frac{k}{4\omega}\left[\frac{2\cos(2(\omega t+\delta)-\tilde{\phi}_2)}{\sqrt{\left(\frac{1}{2}\right)^2+\left(\frac{c}{\omega}\right)^2}}-\frac{1}{2}\frac{\cos(4(\omega t+\delta)-\phi_4)}{\sqrt{1+\left(\frac{c}{\omega}\right)^2}} +\frac{3\omega}{2c}\right]\right)^{1/4}}~,   \quad \tilde{\phi}_2=\frac{\pi}{2}+{\rm arctan}\left(\frac{2c}{\omega}\right)~, \quad \phi_4=\phi_2~.
\end{equation}
Its limit $c\to0$ is smooth after $c_1\to c_1+3k/(8c)$ and recovers \eqref{3-26} with $\delta=\pi/2$.  
	
\begin{figure}[H]
	\centering
    \includegraphics[width=0.78\linewidth]{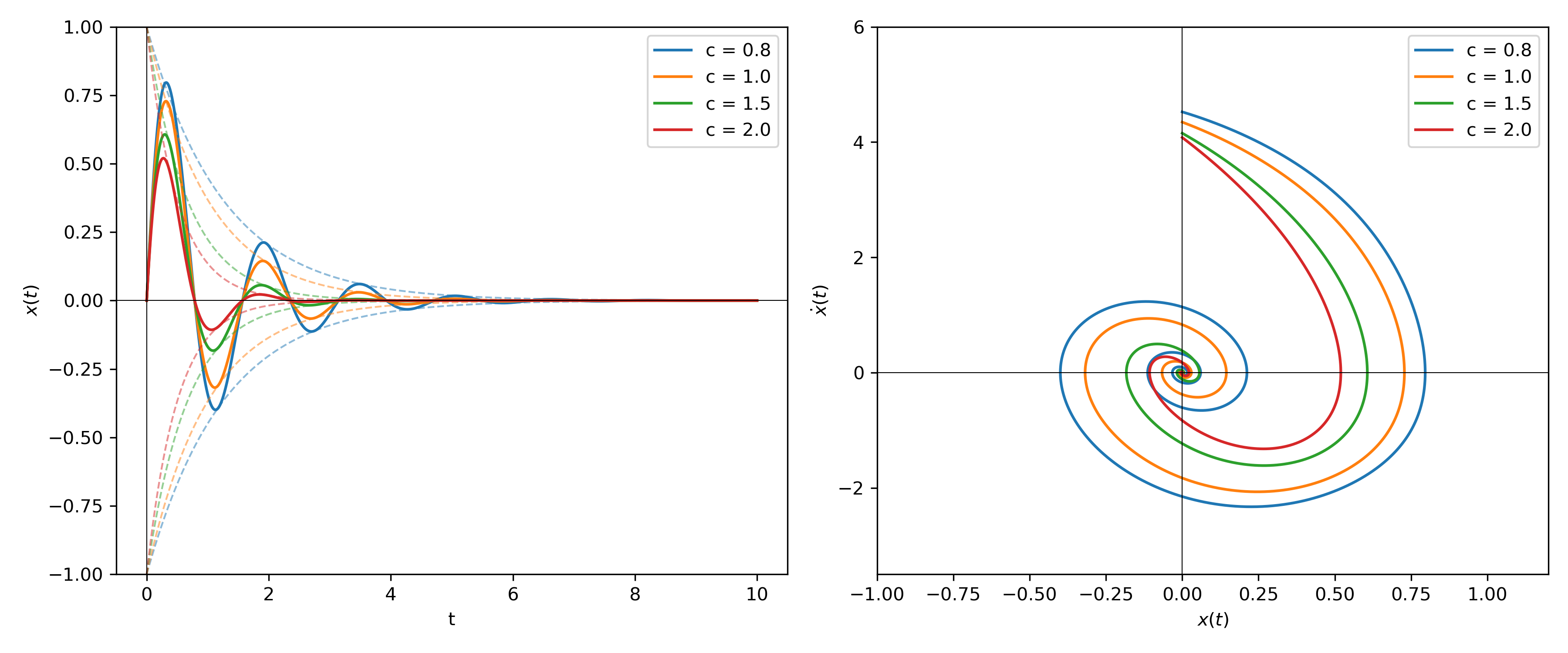}     
	\caption{Case $q=4$ with $\omega=4$, $k=c_1=1$ and $c$ = 0.8, 1.0, 1.5, 2.0. Left: solution $x_4(t)$ with smooth exponential decay governed by $c$. Right: phase portrait showing spirals converging to the origin.}
	\label{fig:q4}
\end{figure}

	Figure \ref{fig:q4} shows the underdamped behavior and the phase portrait for the same parameters as in the previous cases.
	
{\bf Singular ranges of $c_1$}

Writing the denominators of \eqref{eq:q1_ud}--\eqref{eq:q4} as $\Delta_q(t)^{1/q}$, with $\Delta_q(t)=c_1e^{qct}-B_q(t)$ and $B_q$ the subtracted term in each formula, one checks that
\begin{equation}\label{eq:Delta}
\dot\Delta_q=qc\,\Delta_q+qk\sin^q(\omega t+\delta)~.
\end{equation}

\medskip

For even $q$, $\sin^q\ge0$ gives $\Delta_q(t)\ge\Delta_q(0)\,e^{qct}$, so the solution is real and regular for all $t\ge0$ if and only if $\Delta_q(0)>0$, i.e.\ $c_1>B_q(0)$. For $\delta=0$,
\begin{equation}\label{eq:eventhr}
q=2:\ \ c_1>\frac{k\,\omega^2}{2c(\omega^2+c^2)}~,\qquad
q=4:\ \ c_1>k\left[\frac{3}{8c}-\frac{2c}{\omega^2+4c^2}+\frac{c}{8(\omega^2+c^2)}\right]~.
\end{equation}
Both thresholds grow like $1/c$ as $c\to0$, in agreement with the redefinitions of $c_1$ for each case. 
For odd $q$, $B_q$ is purely oscillatory and takes the values $\pm A_q$, $A_q=\max_t|B_q|$, in every period $2\pi/\omega$. The solution is therefore regular for all $t\ge0$ if $|c_1|>A_q$, and it has a singularity in the first period if $|c_1|\le A_q\,e^{-2\pi qc/\omega}$; in between, the outcome depends on $\delta$. One has $A_1=k/\sqrt{\omega^2+c^2}$ and
\begin{equation}\label{eq:A3}
\frac k4\left|\frac{9}{\sqrt{\omega^2+9c^2}}-\frac{1}{\sqrt{\omega^2+c^2}}\right|\le A_3\le\frac k4\left(\frac{9}{\sqrt{\omega^2+9c^2}}+\frac{1}{\sqrt{\omega^2+c^2}}\right)~,
\end{equation}
and, as $c\to0$, $A_1\to k/\omega$ and $A_3\to 2k/\omega$, the isochronous bounds of Section IV.

\subsection{The case $c<0$ and a three-regime classification}

For $c<0$ the origin is unstable, since the eigenvalues $-c\pm i\omega$ have positive real part. By \eqref{eq:rhoc}, asymptotically $\rho_c^{-q}\to\frac{qk}{\omega^q}P(t)$, where $P$ is the periodic solution of $\dot P+q|c|P=\cos^q(\theta_0-\omega t)$. For even $q$, $\cos^q\ge0$ implies $P>0$, and all solutions approach an explicit limit cycle of period $2\pi/\omega$, as already noticed in CSKL for the corresponding equation (30). For odd $q$, $\cos^q$ has zero mean, so $P$ changes sign and $\rho_c$ blows up in finite time. Together with the cases $c=0$ and $c>0$, this gives the following classification for $k>0$:

\medskip

\begin{itemize}

\item $c=0$
\begin{itemize}
\item odd $q$ $\qquad$  isochronous centre (singular $c_1$-range).
\item even $q$ $\qquad$ decay $\sim t^{-1/q}$.
\end{itemize}

\item $c>0$
\begin{itemize}
\item odd $q$ $\qquad$  exponential decay $\sim e^{-ct}$ (singular range).
\item even $q$ $\qquad$  exponential decay $\sim e^{-ct}$.
\end{itemize}

\item $c<0$
\begin{itemize}
\item odd $q$ $\qquad$  finite time blow-up.
\item even $q$ $\qquad$ explicit limit cycle.
\end{itemize}
\end{itemize}

\renewcommand\thesection{\arabic{section}}
	\section{Possible applications}\label{sec10}  
Beyond its role as a rare exactly solvable nonlinear oscillator --- itself the main reason
this family is repeatedly used in quantization schemes and nonlinear-dynamics
methods \cite{bagchi2025,Mustafa2023} --- the $q=1$ case is connected
to position-dependent-mass (PDM) quantum mechanics. Its non-standard Hamiltonian may be considered close to
the operator-ordering resolution (BenDaniel--Duke, Zhu--Kroemer, Li--Kuhn,
Mustafa--Mazharimousavi) that was originally developed to fix the effective-mass ordering
ambiguity at graded and abrupt III--V semiconductor heterojunctions \cite{RegoMonteiro2016}.
The oscillator could be best understood not as a model of a
specific device, but as a solvable calibration case for a PDM quantization resource 
that is independently applied to real graded-composition semiconductor
heterostructures and quantum wells. We also notice as a more specific connection 
that the quantum counterpart of the modified Emden family --- like the
semiconductor-motivated PDM Hamiltonian of \cite{RegoMonteiro2016} itself --- has been
studied as a $\mathcal{PT}$-symmetric system \cite{Ruby2012}.

The underdamped even-$q$ branch connects, in turn, to mechanical vibrational resonance. A
particle constrained to a rotating parabolic wire obeys a damped, position-dependent-mass
equation that reduces, by suitable transformation, to a quintic oscillator --- the same
nonlinearity order as the $q=2$ member of (\ref{eq0}) --- and vibrational resonance has been
studied directly in this system as well as in the closely related damped
Mathews--Lakshmanan oscillator \cite{Kabilan2023epjp,Kabilan2023jvet}. 

Very recently, in pump--probe experiments on the two-dimensional magnetic semiconductor CrSBr, Brennan {\em et al} \cite{Brennan2026} modelled the asymmetric sawtooth waveforms of the magnon modes driven by excitonic spin torque with the Li\'enard equation
\begin{equation}\label{eq:magnon}
X''+(\gamma-\mu X)X'+\frac{\mu^2}{9}X^3+\omega_0^2X=0~,
\end{equation}
where $X$ is the angular displacement of the magnetization. For $\gamma=0$, \eqref{eq:magnon} is the $q=1$ modified Emden oscillator \eqref{eq0} with $k=-\mu/3$ and $\omega=\omega_0$, whose exact solution \eqref{3-18b} is periodic and markedly asymmetric when $|c_1|$ approaches $|k|/\omega$. For $\gamma\neq0$, \eqref{eq:magnon} differs from the $q=1$ member of the integrable family \eqref{eq:underdamped_lienard}, with $c=\gamma/2$ and $\omega_0^2=\omega^2+\gamma^2/4$, only by the absence of the quadratic term $-\mu\gamma X^2/3$; neither a shift of $X$ nor an autonomous Sundman map brings it into the integrable class. The exact waveforms of Section IX thus provide a natural reference for the damped experimental signal. 

\renewcommand\thesection{\arabic{section}}
	\section{Conclusions}\label{sec11} 

We have summarized the relation of the results obtained here to earlier work on modified Emden oscillators at the end of the Introduction.

The waveforms of the modified Emden oscillators of arbitrary natural power $q$, and of their damped $c$-deformed versions, have been obtained in closed form through equivalent Bernoulli equations generated by a commutative factorization. We have shown that this factorization is not an independent method but a gauge-conjugated form of the nonlocal CSKL transformation: a Li\'enard equation admits a commutative factorization if and only if it is the image of a constant-coefficient linear oscillator under the CSKL map, the factorization constant being half the difference of the linear roots. Its advantage is algebraic: the Riccati equation of the CSKL approach is replaced directly by a Bernoulli equation.

The dynamics has a unique equilibrium at the origin for every $q$, and a clear odd--even dichotomy. For odd $q$ all bounded solutions are isochronous, with period $T=2\pi/\omega$ independent of the amplitude, and the admissible range of the integration constant is given explicitly; the forbidden range grows like $\sqrt q$. For even $q$ the secular term in the denominator produces algebraic decay, $x\sim t^{-1/q}$, and stable spirals in the phase plane. The dichotomy is confirmed geometrically by the elliptic polar representation, in which the phase rotates uniformly for all $q$ while the radial velocity is sign-definite only for even $q$, and, for even $q$, by the Bendixson--Dulac criterion.

For every $q$, the oscillators admit a Lagrangian with a position-dependent Lurie-Rayleigh function, which is a non-negative dissipation function only for even $q$ and $k>0$; for odd $q$ it describes balanced gain and loss, whose integral over each isochronous orbit vanishes.

For $q=1$, an autonomous Sundman linearization gives a rational first integral, a Darboux factor, a Jacobi last multiplier and a non-dissipative Lagrangian, the latter coinciding with that of Chandrasekar, Senthilvelan and Lakshmanan. For every $q$, $\omega^2x^2+(\dot x+kx^{q+1})^2$ is a Darboux polynomial and its power $-(q+2)/2$ a global Jacobi last multiplier, while a global (rational) first integral exists only for odd $q$.

Introducing a real damping constant $c>0$ in the factorization yields exponentially decaying waveforms with decay rate $c$ independent of $q$, while $c<0$ leads to explicit limit cycles of period $2\pi/\omega$ for even $q$ and to finite-time blow-up for odd $q$. Finally, the $q=1$ member of the family appears in recent models of magnon dynamics driven by excitonic spin torque in CrSBr, where the asymmetric waveforms of the exact solutions provide a natural reference for the experimental signals.

    	\section*{Acknowledgement} 
    The first author acknowledges the financial support of SECIHTI through a postdoctoral fellowship. Both authors are grateful to the referees for their very helpful suggestions which greatly improved this paper. 
    
    \medskip

\appendix
\setcounter{equation}{0}
\def\theequation{A\arabic{equation}}
\section{Comparison with a class of Gin\'e--Sinelshchikov nonlinear oscillators} 
%
Example~2 of \cite{GS2024} is the family of nonlinear oscillators
\begin{equation}\label{A1}  
z_{\xi\xi}+(\mu+z^\frak{m})z_\xi+\frac{2}{9(\frak{m}+1)}z^{2\frak{m}+1}+\frac{\mu}{\frak{m}+2}z^{\frak{m}+1}+(\frak{m}+1)\frac{\mu^2}{(\frak{m}+2)^2}z=0~, 
\end{equation}
which the authors map to the anharmonic oscillator $y_{{\rm x}{\rm x}}+(\frak{p}+1)\tilde{\delta}y^\frak{p}=0$ with $\frak{p}=(1-\frak{m})/(3\frak{m}+1)$ and $\tilde{\delta}=-1$.
Comparison with the $\lambda$-modified Emden equation $\ddot x+3kx\dot x+k^2x^3+\lambda x=0$ with its damping linear in $x$ shows that only $\frak{m}=1$ can match, which gives
\begin{equation}\label{A2} 
z''+(z+\mu)z'+\tfrac19z^3+\tfrac{\mu}{3}z^2+\tfrac{2\mu^2}{9}z=0~.
\end{equation}
Matching further these two equations requires  $k=1/3$ and $\mu=0$, hence $\lambda=0$. On the other hand, performing the shift $x=z+\mu$ into the $\lambda$-modified Emden equation with $k=1/3$ gives
\[
\ddot z+(z+\mu)\dot z+\tfrac19z^3+\tfrac{\mu}{3}z^2+\Big(\tfrac{\mu^2}{3}+\lambda\Big)z+\mu\Big(\tfrac{\mu^2}{9}+\lambda\Big)=0~,
\]
and $\lambda=-\mu^2/9$ removes the constant term while turning the linear coefficient into $2\mu^2/9$. Thus \eqref{A2} is the  $\lambda$-modified Emden
equation  with $k=1/3$, $\lambda=-\mu^2/9$, $x=z+\mu$. Since $\lambda<0$ for $\mu\ne0$, it does not attain the isochronous regime $\lambda>0$. Equivalently, \eqref{A2} is the generalized modified Emden equation CSKL (9) with $k=1/3$, $\frak{c}_1=\mu$, $\frak{c}_2=2\mu^2/9$. The underlying linear oscillator has roots $-\mu/3$ and $-2\mu/3$.

More generally, \eqref{A1} 
belongs to the autonomous CSKL family, i.e.\ CSKL (13) with $\frak{n}=1$ and constant $\gamma$, if and only if $\frak{m}=1$ or $\frak{m}=-\frac12$, for every $\mu$. Indeed, CSKL (13) then reads $\ddot x+\big(\frak{c}_1+2\gamma+(\frak{m}+2)\beta x^\frak{m}\big)\dot x+\beta^2x^{2\frak{m}+1}+\beta(\frak{c}_1+2\gamma)x^{\frak{m}+1}+(\gamma^2+\frak{c}_1\gamma+\frak{c}_2)x=0$; taking $\beta=1/(\frak{m}+2)$ and $\frak{c}_1+2\gamma=\mu$ matches the damping and, automatically, the $x^{\frak{m}+1}$ term, while the linear term is matched by $\frak{c}_2$.

The only remaining condition is $1/(\frak{m}+2)^2=2/(9(\frak{m}+1))$,
i.e.\ $2\frak{m}^2-\frak{m}-1=0$.
The same condition follows from a direct term-by-term comparison with the
$c$-deformed equation~\eqref{eq:underdamped_lienard}: with $x=z$,
$q=\frak{m}$, $k=1/(\frak{m}+2)$ and $c=\mu/2$, the coefficient of
$x^{2q+1}$ again requires $2\frak{m}^2-\frak{m}-1=0$, the $x^{q+1}$ term is
matched automatically, and the linear term gives $\omega^2=-\mu^2/36$ for
both roots. Hence \eqref{A1} is the overdamped continuation
$\omega=i\mu/6$ (i.e.\ $c=3|\omega|$) of the $c$-deformed oscillator, whose
linear roots $-c\pm i\omega$ are precisely $-\mu/3$ and $-2\mu/3$.
For $\frak{m}=1$, i.e.\ $q=1$, its general solution therefore follows from
\eqref{eq:gensol} with $\sin(\omega t)\to i\sinh(\mu t/6)$ (and $c_1$
redefined accordingly), whereas $\frak{m}=-\frac12$ lies outside the range
of positive integer $q$ considered here.
These identifications must not be confused with the exponent $\frak{p}$,
which refers to a different equation: $q=\frak{m}$ is obtained from
\eqref{A1} with no change of variables, whereas
$\frak{p}=(1-\frak{m})/(3\frak{m}+1)$ is the exponent of the anharmonic
oscillator reached through the point transformation (4.22) of
\cite{GS2024}, which changes both $z$ and $\xi$.
Thus $\frak{m}=1$ corresponds to $\frak{p}=0$, i.e.\ to the trivially
linear target $y_{\rm xx}=1$ (for $\tilde\delta=-1$): this is the same
linearizable case seen from the other side, and it is excluded as trivial
by the hypotheses of Theorem~3 of \cite{GS2024}.
On the other hand, $\frak{m}=-\frac12$ corresponds to $\frak{p}=-3$.
Since $\tilde\delta$ is a constant, equal to $-1$ in Example~2 of
\cite{GS2024}, the target equation is then $y_{\rm xx}+2y^{-3}=0$, i.e.\
the Ermakov--Pinney equation $y''+\Omega^2y=\lambda y^{-3}$ with $\Omega=0$
and attractive coupling $\lambda=2\tilde\delta=-2$.
The same two values, $\frak{m}=1$ and $\frak{m}=-\frac12$, are exactly
those for which \eqref{A1} satisfies the autonomous Sundman
linearization condition~\eqref{eq:Sinh}, taken from Corollary~2.2 of
\cite{Sin2020}, with $\alpha/\tilde\beta^2=2/9$ as in~\eqref{8.9}.

\medskip
\bigskip

{\bf Data availability statement}\\

No new data were created or analysed in this study.


\end{document}